\documentclass[11pt]{article}
\usepackage[english]{babel}
\usepackage[utf8]{inputenc}

\usepackage{calc}
\usepackage{amsmath}
\usepackage{graphicx}
\usepackage{epstopdf}
\usepackage{dirtytalk}
\usepackage{amsfonts}

\usepackage{pgfgantt}
\usepackage{xcolor}
\usepackage{advdate}
\usepackage{comment}
\usepackage{booktabs}

\usepackage{url}
\usepackage{cite}

\usepackage{tikz}
\usetikzlibrary{patterns,patterns.meta}
\usetikzlibrary{shapes}
\usetikzlibrary{arrows.meta}
\usetikzlibrary{calc,shadows,matrix}
\usetikzlibrary{backgrounds}
\usetikzlibrary{decorations.pathreplacing}
\usetikzlibrary{positioning}

\usepackage{tikzscale}

\usepackage[numbers,sort&compress]{natbib}
\usepackage{pgfplots}
\usepgfplotslibrary{patchplots}
\usepgfplotslibrary{fillbetween}
\usepgfplotslibrary{polar} 
\usetikzlibrary{pgfplots.polar}
\usepackage[top=2cm, bottom=2cm, right=2cm, left=2cm]{geometry}

\usepackage[left]{lineno} % place numbers on the left
\usepackage{siunitx}
\newcommand{\keywords}[1]{\vspace{1em}\noindent\textbf{Keywords:} #1}

\title{Matching frequency response measurements and reduced order models for the inverse identification of viscoelastic properties}

\author{Linus Taenzer$^{a}$$^{b}$$^{*}$, Paolo Tiso$^{b}$, Bart Van Damme$^{a}$ \\\\
        \small $^{a}$ Laboratory of Acoustics/ Noise Control, Empa, Überlandstrasse 129, 8600 Dübendorf, Switzerland \\
        \small $^{b}$ Institute for Mechanical Systems, ETH Zürich, Leonhardstraße 21, 8092 Zürich, Switzerland \\
        \small $^{*}$Corresponding author: Linus Taenzer; \tt{linus.taenzer@empa.ch} \\
}
\date{}
\pgfplotsset{compat=1.17} 
\begin{document}

\maketitle
%\linenumbers 
\begin{abstract} 
\noindent 
3D-printed materials are used in many different industries (automotive, aviation, medicine, etc.). Most of these 3D-printed materials are based on ceramics or polymers whose mechanical properties vary with frequency. For numerical modeling, it is crucial to characterize this frequency dependency accurately to enable realistic finite-element simulations. At the same time, the damping behavior plays a key role in product development, since it governs a component’s response at resonance and thus impacts both performance and longevity. In current research, inverse material characterization methods are getting more and more popular. However, their practical validation and applicability on real measurement data have not yet been discussed widely. In this work, we show the identification of two different materials, POM and additively manufactured sintered ceramics, and validate it with experimental data of a well-established measurement technique (dynamic mechanical analysis). The material identification process considers state-of-the-art reduced-order modeling and constrained particle swarm optimization, which are used to fit the frequency response functions of point measurements obtained by a laser Doppler vibrometer. This work shows the quality of the method in identifying the parameters defining the viscoelastic fractional derivative model, including their uncertainty. It also illustrates the applicability of this identification method in the presence of practical difficulties that come along with experimental data such as boundary conditions and noise.

\end{abstract}

\noindent \keywords{Viscoelasticity, model order reduction, vibrations, inverse identification, dynamical mechanical analysis}\\

\section{Introduction}

Additively manufactured structures made of polymers and ceramics play an increasing role in a wide range of applications, from aerospace and automotive components (vibration damping, impact protection, sealing, gaskets, etc.) to biomedical devices (prostheses, implants) and electronics (wearable devices), thanks to their design freedom, simple integration that allows for custom products, and fast development processes. Depending on the application, the mechanical, thermal, and chemical properties affect their compatibility~\cite{Bose2024,Saleh2021,ceramics7010006}. 
One of the mechanical properties that influences the dynamic performance and durability, but is not yet investigated in a rigorous manner, is viscoelasticity, which describes the time-dependent deformation under stress. In contrast to purely elastic materials, viscoelastic materials exhibit both elastic and viscous behavior, which means that they store and dissipate mechanical energy depending on the duration and speed of the applied forces. Viscoelastic behavior can be described in either the time or frequency domain. The former is suitable to analyze the response to constant and transient loads, while the latter is appropriate for responses to cyclic excitation~\cite{ROULEAU2017384,MASTRODDI201942,li2021investigation}. Several authors have shown that an elastic material model is not sufficient for e.g. predicting band-gaps in metamaterials~\cite{Krushynska2021,Krushynska2023,Moleron_2016} or accurately modeling the stiffness since the printing orientation affects the glass transition temperature and thereby the stiffness for different printing directions~\cite{Kontaxis2025}. 

Traditionally, viscoelastic properties in the frequency domain are determined using experimental methods such as dynamic mechanical analysis (DMA), where materials are subjected to oscillatory stresses at different frequencies and temperatures, and the corresponding strain response is measured~\cite{menard2008dynamic,baz2019active,Yi2023}. The material’s storage modulus (elastic behavior) and loss modulus (viscous behavior) are derived from these measurements, providing insight into how the material stores and dissipates mechanical energy at different frequencies. However, these experimental techniques have limitations, including the need for multiple tests over a wide range of frequencies and environmental conditions, which can be time-consuming. The tests are done on small samples (typical length in the order of a few centimeters and 1-3 mm thickness) and their size and mounting are sources of measurement errors~\cite{polym12081700}. Furthermore, it has been shown that these results can not be generalized for every geometry and that the geometric shape might require form factor compensation~\cite{CLAMROTH1981263,SEOUDI2009495}. The resulting master curves are therefore not always accurately describing the intricate behavior of large complex shapes under combined compression and shear loading. 

Inverse identification schemes offer a popular alternative. They essentially consist of optimizing material properties by minimizing, in some sense, the distance of a numerical response (typically retrieved from finite element (FE) simulations) from experimental data. Different loss functions have been considered to identify viscoelastic parameters from frequency response functions~\cite{HAMDAOUI2019237}. Typically, simple structures such as beams or plates of the unknown material are considered since the models are computationally beneficial~\cite{KIM2009570, SHI20061234,MARTINEZAGIRRE20113930,NGO2018172}. Still, for large frequency ranges and in case of a large space of  material properties, evaluating a numerical model of such a structure can be computationally expensive. Many of these methods struggle to balance computational efficiency due to large matrix factorizations with the need for precision across a wide frequency range, limiting their practicality in real-world applications. \citet{SHI20061234} note that their inverse finite-element-based identification procedure requires iterative updates of material parameters to match experimental resonance frequencies, and that poor initialization can lead to excessive computation time and convergence issues. \citet{KIM2009570} acknowledge that conventional FRF-based identification is time-consuming and that their iterative FE-based optimization remains computationally demanding due to repeated sensitivity analyses. Similarly, \citet{MARTINEZAGIRRE20113930} highlight accuracy degradation in modal-based approaches for highly damped materials due to modal overlap and strong frequency dependence.

Recent advances in model order reduction (MOR) offer new opportunities to facilitate the identification of viscoelastic properties in the frequency domain. In fact, the optimization process involved in identifying material parameters to match precise experimental data can greatly benefit from the accelerated solutions provided by reduced-order models. These reduced-order models are generated via a Krylov subspace approach, specifically employing a second-order Arnoldi algorithm. However, to the best of our knowledge they have only been validated on numerical data with added artificial noise rather than actual experimental measurement data, leaving questions about robustness and generalization~\cite{Zhang2023,Xie2019,VandeWalle}. These authors have been working on identifying frequency-dependent material properties using MOR, and different viscoelastic material models. Additionally modal projections techniques have been compared among themselves for a known frequency-dependent material targeting the accuracy of the dynamic response~\cite{ROULEAU2017110}. In their work, Rouleau et al. do not include an automated procedure to identify those constitutive parameters. For the fast computation of the dynamic response of viscoelastic structures, a framework using parametric model order reduction (pMOR) was introduced by~\citet{Xie2018} using Krylov subspaces and a Design Of Experiment (DOE) to assemble multiple frequency-dependent bases. This framework was used in an inverse sense to characterize the material properties in a numerical experiment~\cite{Xie2019}. Alternatively,~\citet{Aumann2022} developed an automatic method using a rational form for approximating the frequency dependency and perform the MOR using the adaptive Antoulas Anderson algorithm (AAA). This automation, however, applies to the frequency-domain reduction step rather than to a full parametric model-order-reduction framework.
Dynamical systems stemming from Newton's law (as for the case we are treating here) are of second order. They can be transformed to first order state-space form to comply with the large bulk of dynamical system theory. This is done at the cost of doubling the size of the unknowns, which now feature also the generalized velocities alongside the generalized displacements. Because a Krylov subspace is constructed from the system’s input vector, it naturally suits forced vibration problems. Moreover, it enables the incorporation of frequency-dependent stiffness variations, which enhances the accuracy of the reduced model--something that stands in contrast to modal methods, where strongly frequency-dependent properties cannot be easily accommodated within eigenvector-based approaches. For reduction, the Arnoldi algorithm is adopted~\cite{arnoldi1951minimized}. It provides a reduction basis for both velocities and displacements. However, its vanilla application poses computational difficulties such as memory use, worse conditioning, and convergence issues due to numerical stability. Because of this, the second order Arnoldi Algorithm (SOAR) was proposed~\cite{Bai2005}. This improved version is computationally parsimonious by preserving the structure  of the equations, thus allowing the extraction of a reduction basis for many expansion points. With the two-level orthogonal Arnoldi algorithm (TOAR), the stability and efficiency of finding a subspace was further improved \cite{Lu2016}. 
 In this work, we show the applicability of the discussed pMOR algorithms, and explore which optimization methods are best suited for improving the material identification process based on experimental data from 3D-printed structures with complex shapes. This is an important novelty compared to previous studies of a purely numerical nature, making the proposed technique applicable for a broader user field. The work includes an overview of possible errors coming from small deviations of the measurement position and the choice of the fitted transfer functions. We review the investigated adaptive MOR method developed by~\citet{Xie2019}, which is used to assess its applicability for polymer and ceramic structures. In the sections ~\ref{sec:visco} and ~\ref{sec:numerical_modelling}, the theory of viscoelastic materials and MOR is summarized. We then introduce the setup for reliable measurements obtained by using a Scanning Laser Doppler Vibrometer (SLDV), which measures the normal velocities on a structure's surface, shown in section~\ref{sec:opt}. The material properties are optimized using the obtained measurements of FRFs and an adapted MOR framework, incorporating particle swarm optimization. The results are presented for two distinct use cases. The first use case compares the proposed method to DMA measurements for a standard homogeneous polymer called polyoxymethylene (POM), showcasing advantages in efficiency, accuracy, and applicability to numerical models~\cite{menard2008dynamic}. The second use case focuses on more complex composite materials of 3D-printed polymer-ceramic samples having a curved structure.

\section{Viscoelastic material model}
\label{sec:visco}
\begin{figure}
    \centering
    \includegraphics[width=0.9\linewidth]{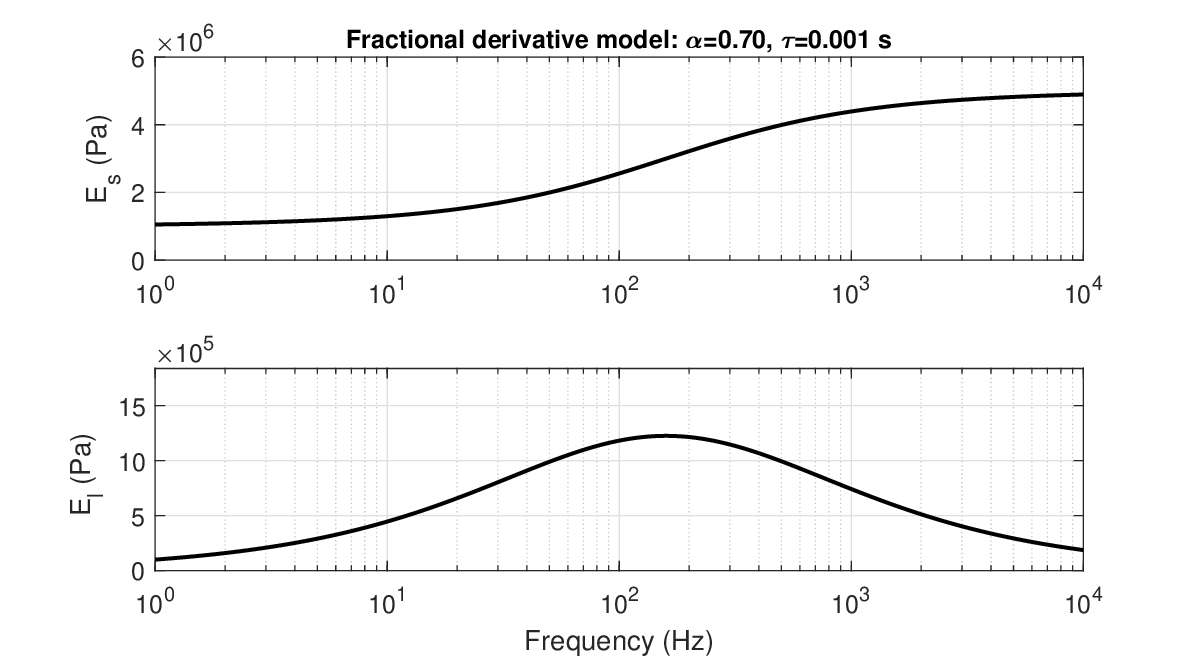}
    \caption{Exemplary material behavior of the fractional derivative model, illustrating the storage modulus $E_s$ and loss modulus $E_l$.}
    \label{fig:visco_curve}
\end{figure}

Viscoelasticity describes the properties of a material that has both viscous and elastic behavior. In the frequency domain the energy dissipation during cyclic loading can be approximated by the complex Young's modulus 
\begin{equation}\label{eq:complexmod}
    E(s) = E_s(s)+i E_l(s),
\end{equation}
where $E_s$ and $E_l$  are the storage and loss modulus respectively. Depending on the testing procedure, the complex shear modulus $G(s)$ can also be considered, in this work we will however consistently use the Young's modulus. In steady state cases,  both moduli depend on the angular frequency of excitation $\omega$, or equivalently the complex angular frequency $s=i \omega$. A commonly used derived metric for moderate damping is the loss factor, 
\begin{equation}
    \tan \delta(s) = \frac{E_l(s)}{E_s(s)}
\end{equation}
describing the magnitude of damping at each frequency by a phase difference $\delta(s)$.
The Maxwell and Kelvin-Voigt models are commonly utilized as lumped parameter models to describe the viscoelastic behavior of materials. These models represent complex material properties through discrete mechanical elements, such as springs and dash-pots, which capture elastic and viscous responses, respectively.  These simplified mechanical analogs reduce the continuous distribution of material properties into a finite number of elements, simplifying the analysis of time-dependent features like creep and stress relaxation~\cite{Basz2019}. Although these models are widely used, they cannot quantitatively reproduce the more complex dynamic response of industrial materials. Often the modeled curves rise more steeply than those observed experimentally which makes identification schemes challenging. To reproduce frequency-dependent dynamic properties more accurately, more complex models such as the generalized Maxwell model or Gollah-Huges-McTavish model have been proposed~\cite{Basz2019}. However these models require a large, a priory unknown, number of fitting parameters. As a compromise, fractional models have been introduced to allow a good agreement with physical material parameters~\cite{Koeller1984}. 
In the 1980s,~\citet{Bagley1983} presented the fractional derivative model which is equivalent to a power law relaxation time. The complex modulus is given by 
\begin{equation}
    E(s) = \frac{E_0 + (s \tau)^{\alpha}E_{\infty}}{1+(s \tau)^{\alpha}},
\end{equation}
where $E_0$ is the static modulus, $E_\infty$  the high frequency limit modulus, $\tau$ the relaxation time, and $\alpha$ the fractional order coefficient~\cite{PRITZ1996103}. For the two extreme cases of the fractional coefficient $\alpha$, this function either reduces to the Maxwell model ($\alpha = 1$) or to a Hookean solid ($\alpha=0$)~\cite{Bagley1983,Koeller1984}. In this work, the fractional derivative model is used, since it can approximate the majority of materials with only four free parameters. This model has been applied for example for capturing DMA data over a wide range of frequencies and temperatures~\cite{Rouleau2015,Pawlak2021}. Looking at the full frequency range, $E_0$ and $E_{\infty}$ provide the lower and upper bound of the storage modulus, whereas the relaxation time defines the frequency of maximum damping as well as the inflection point in the storage modulus curve. This is illustrated using a dummy material model in Fig.~\ref{fig:visco_curve}. The fractional coefficient $\alpha$ narrows or widens this transition range.

%In the following, the dependence on $s$ will be omitted from the notation for simplicity.

\section{Numerical modeling of transfer functions in viscoelastic structures}
\label{sec:numerical_modelling}
\subsection{Finite element formulation}
Upon FE discretizations, the forced vibration problem with a frequency-dependent viscoelastic material model $E(s)$ can be written as
\begin{equation}\label{eq:eom}
\left(E(s)\mathbf{K}_v + s^2\mathbf{M}\right) \mathbf{u} = \mathbf{f},
\end{equation}
where $\mathbf{K}_v$ is the stiffness matrix for unit Young's modulus of the viscoelastic domain and $\mathbf{M}$ is the mass matrix. The system is excited by the spatially distributed harmonic input force vector $\mathbf{f}$ and solved for the displacement $\mathbf{u}\in \mathbb{R}^n$, where $n$ is the size of the full order model. For a given complex and frequency-dependent Young's modulus, combining eq.~(\ref{eq:complexmod}) and eq.~(\ref{eq:eom}) yields the classical form of the discretized damped equation of motion
\begin{equation}\
\left(E_s(s)\mathbf{K}_v + s \frac{E_l(s)}{\omega}\mathbf{K}_v + s^2\mathbf{M}\right) \mathbf{u} = \mathbf{f}.
\end{equation}
In order to map the transfer functions with respect to a unit force $\|\mathbf{f}\|=1$, the system can be rewritten as
\begin{equation}\label{eq:FRF}
h(s) = \mathbf{l}^T ( s^2 \mathbf{M} + E(s) \mathbf{K}_v)^{-1}\mathbf{f}, 
\end{equation}
with $h(s)$ describing the input-output behavior and $\mathbf{l}$ being the output vector selecting the nodal quantity of interest.

\subsection{MOR with frequency-dependent basis}
%\subsubsection{Krylov subspace generation}
The system of eq.~(\ref{eq:eom}) bears prohibitive computational costs due to the inversion of the large matrix in the left-hand side for each frequency step, and therefore cannot be used for efficient material property identification. A reduced version of this dynamic stiffness matrix is needed. We follow the frequency-dependent MOR algorithm as described in~\cite{Xie2018}, giving a short summary of the basic principles provided in this work. The details on the expansion point selection can be found in the mentioned paper.  

The frequency dependency described by the chosen viscoelastic material model, in this case the fractional derivative model, is approximated by using a Taylor series expansion of second order
\begin{equation}
\bar{E}(s) = E(s_0) + E'(s_0)(s - s_0) + \frac{E''(s_0)}{2}(s - s_0)^2 + R_2(s).
\end{equation}
Expansion points $s_0=i2\pi f_0$ at as many frequencies as needed are selected to keep the solution accurate over the entire investigated frequency range, and for each expansion point a reduced-order model is developed. 
By inserting the expanded viscoelastic model in eq.~(\ref{eq:eom}), the system can be rewritten as 
\begin{equation}
\left( \Tilde{\mathbf{K}} + (s - s_0) \Tilde{\mathbf{D}}  + (s - s_0)^2 \Tilde{\mathbf{M}}  + R_2(s) \Tilde{\mathbf{K}}  \right) \mathbf{u} = \mathbf{f},
\end{equation}
where $\Tilde{\mathbf{K}}  =  E(s_0) \mathbf{K}_v + s_0^2 \mathbf{M}$, $\Tilde{\mathbf{D}}  = 2 s_0 \mathbf{M} + E'(s_0) \mathbf{K}_v$, and $\Tilde{\mathbf{M}}  = \mathbf{M} + \frac{E''(s_0) \mathbf{K}_v}{2}$. It should be pointed out that the choice for a second-order Taylor series maintains the second order in $s$, and is therefore the natural option to approximate the equation of motion in eq.~(\ref{eq:eom}). 
The goal is to build a much smaller model $\hat{h}(s)$ which approximates the original transfer function defined in eq.~(\ref{eq:FRF}) by matching the first $k$ terms $m_k$ (its leading moments) at $s_0$: 
\begin{equation}
    	h(s) \approx \hat{h}(s) =  m_{0} + m_{1}(s-s_0) + m_{2}(s-s_0)^{2} + ... +m_{k-1}(s-s_0)^{k-1}.
\label{eq:pwr_series}
\end{equation} 
These moments can be computed by determining the Krylov subspace vectors around a selected expansion point $f_0$. A Krylov subspace generates a projection based solution space of a dynamic system. For a second order dynamic system the definition of a Krylov subspace is  
\begin{equation}
    \mathcal{K}_{n}(\mathbf{A},\mathbf{B},\mathbf{r}_0) := \text{span}\{\mathbf{r}_0,\mathbf{r}_1,...,\mathbf{r}_{n-1}\},
\end{equation}
where 
$\mathbf{A}=\Tilde{\mathbf{K}}^{-1}\Tilde{\mathbf{D}}$, $\mathbf{B} =-\Tilde{\mathbf{K}}^{-1}\Tilde{\mathbf{M}}$ and $\mathbf{r_0}=\Tilde{\mathbf{K}}^{-1}\mathbf{f}$. 
With the two initial base vectors $\mathbf{r}_0$ and $\mathbf{r}_1 = \mathbf{A} \mathbf{r}_0$, further projection vectors are iteratively defined by the two previous iterations
\begin{equation}
\mathbf{r}_k = \mathbf{A} \mathbf{r}_{k-1} + \mathbf{B} \mathbf{r}_{k-2},\quad k \geq 2,
\end{equation}
where $k$ is the number of basis vectors, which is also known as the second order Krylov sequence. 

To create the reduced orthonormal basis $V_r$, the two-level orthogonal Arnoldi algorithm (TOAR) can be applied on the spanned Krylov subspace~\cite{Bai2005,Lu2016}. The number of base vectors in the Krylov subspaces is then iteratively expanded until the relative error between the new and the previous iteration of the approximated FRF meets a predefined convergence criterion. The reduced-order model is obtained by multiplying the original system matrices with the reduced basis, so that the calculation of the reduced transfer function is finally given by 
\begin{equation}
\label{eq:eom_MOR}
h_{\text{MOR}}(s) = \mathbf{l}_r^{\dag} \left( s^2 \mathbf{M}_r + E(s) \mathbf{K}_r\right)^{-1}\mathbf{f}_r, 
\end{equation}
where  $\mathbf{M}_r = \mathbf{V}_r^{\dag} \mathbf{M} \mathbf{V}_r $, $
\mathbf{K}_r = \mathbf{V}_r^{\dag} \mathbf{K} \mathbf{V}_r$, $\mathbf{l}_r = \mathbf{V}_r^{\dag} \mathbf{l}$, $\mathbf{f}_r = \mathbf{V}_r^{\dag} \mathbf{f}$, and $\dag$ denotes the Hermitian (conjugate transpose) of the subspace.

The reduced-order model is accurate for a relatively small frequency range around the expansion point, when keep the numbers of basis vectors at a minimum level, which is needed for fast basis calculation. Therefore an automatic expansion point selection is presented in~\citet{Xie2018}, which is based on a relative error criterion. By creating several bases at different expansion points, a global basis can be assembled using singular value decomposition (SVD). Consider a set of subspaces \({V}_1, {V}_2, \dots, {V}_N\) at expansion points \(f_1, f_2, \dots, f_N\), we assemble these bases into a single subspace
\begin{equation}
\mathcal{V}_{\cup} = \begin{bmatrix}{V}_1 & {V}_2 & \dots & {V}_N \end{bmatrix}.
\end{equation}
On this appended basis $\mathcal{V}_{\cup}$ a singular value decomposition is performed, where
\begin{equation}
\mathcal{V}_{\cup} = \mathbf{U} \Sigma \mathbf{W}^T 
\end{equation}
and \(diag(\Sigma)= (\sigma_1, \sigma_2, \dots, \sigma_{rn})\) contains the singular values, $\mathbf{U}$
is the matrix of left singular vectors and 
$\mathbf{W}$ is the matrix of right singular vectors. We retain only the singular values \(\sigma_i\) such that
\begin{equation}
\sigma_i \geq 10^{-5}  \sigma_{\text{max} \quad i = 1,\dots,r},
\end{equation}
removing numerically insignificant modes while preserving the dominant system dynamics. As shown later, this threshold has proven to be robust for automatically truncating multiple bases while yielding good approximations for the large parameter space of the viscoelastic material model and frequency ranges. 
The common new basis, denoted as \(\mathcal{Q}\), is then formed by the left singular vectors corresponding to the retained singular values
\begin{equation}
\mathcal{Q} = \begin{bmatrix} \mathbf{u}_1 & \mathbf{u}_2 & \dots \mathbf{u}_r \end{bmatrix}.
\end{equation}
By replacing basis $V$ with $Q$ in eq.~(\ref{eq:eom_MOR}) the resulting ROM, valid for a wide frequency interval, is obtained.

\subsubsection{Parametric MOR: design of experiment and basis assembly}
Since the original goal is to inversely identify a material without  prior knowledge, the reduced-order model must not only be accurate for the frequency-dependent properties of a known material but also cover a wide range of parameters to span the material domain.
Therefore, the possible material design space needs to be sampled by a sufficient amount of design points. This number varies depending on the size of the search range of the material model, the frequency range of interest, the structure's modal density, and the number of degrees of freedom (DOF) considered in the reduced-order model. 
The four investigated parameters defining the fractional derivative model are first transformed so that they have similar orders of magnitude

\begin{equation}
\begin{aligned}
    p_1 &= \frac{1}{10} \log_{10}(E_0) \\
    p_2 &= \frac{1}{10} \log_{10}(E_{\infty}) \\
    p_3 &= -\frac{1}{10} \log_{10}(\tau) \\
    p_4 &= \alpha
\end{aligned}
\end{equation}

Based on these transformed parameters, the lower and upper bounds for the DOE are chosen as
\begin{align}
    \mathbf{p}^{\text{lb}} &= \left[ p_1^{\text{lb}}, p_2^{\text{lb}}, p_3^{\text{lb}}, p_4^{\text{lb}} \right] \\
    \mathbf{p}^{\text{ub}} &= \left[ p_1^{\text{ub}}, p_2^{\text{ub}}, p_3^{\text{ub}}, p_4^{\text{ub}} \right].
\end{align}
For the applications in this work, material properties are investigated that lie within the parameter bounds expressed in Tab.~\ref{tab:parameter_bounds}. For many materials there is a nominal stiffness value available in the datasheet but it might differ significantly from $E_0$ and $E_\infty$. For 3D printed structures the nominal data can vary since printing speed, orientation and temperature have a significant impact~\cite{Kontaxis2025}. The inverse optimization based on actual measured data introduces an additional difficulty. Unlike previous numerical studies~\cite{Xie2019}, the experimental data might not always lead to a perfect fit even for noisy data. Imperfections could lead to wrong optima and therefore well-chosen bounds of material data become necessary.  

\begin{figure}[h!]
    \centering
    \begin{minipage}{0.4\textwidth}  % Image takes 60% of the width
        \centering
        \begin{tikzpicture}
            % Include the image
            \node[anchor=south west,inner sep=0] (image) at (0,0) 
            {\includegraphics[width=\textwidth,trim=60 0 60 0, clip]{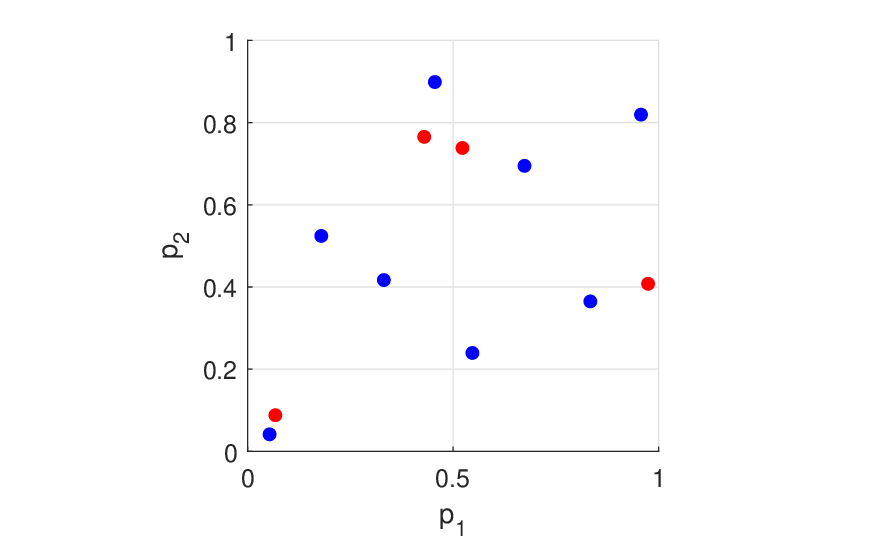}};
        \end{tikzpicture}
    \end{minipage}%
    \hfill
    \begin{minipage}{0.6\textwidth}  % Table takes 35% of the width
        \centering
        \renewcommand{\arraystretch}{1.2}  % Increase row height
        \begin{tabular}{c c c c}
            \toprule
            \( E_0 \) (GPa) & \( E_{\infty} \) (GPa) & \( \tau \) (s) & \( \alpha \) \\
            \midrule
            0.1  & 5.2  & \( 3.6 \times 10^{-4} \) & 0.74 \\
            2.7  & 5.5  & \( 1.7 \times 10^{-7} \) & 0.47 \\
            0.2  & 0.4  & \( 3.8 \times 10^{-1} \) & 0.02 \\
            1.2  & 24   & \( 1.1 \times 10^{-6} \) & 0.75 \\
            \bottomrule
        \end{tabular}
    \end{minipage}
    \caption{Left: DOE (normalized) illustration for quasi-random Sobol sampling points (blue) and validation points (red). Using the global basis, each validation point is compared to the full order model. Right: The material properties of the  corresponding material table of the validation points.}
    \label{fig:image_with_table}
\end{figure}

\begin{figure}[h!]
    \centering
    \begin{tikzpicture}
        % Include the image
        \node[anchor=south west,inner sep=0] (image) at (0,0) {\includegraphics[width=18cm]{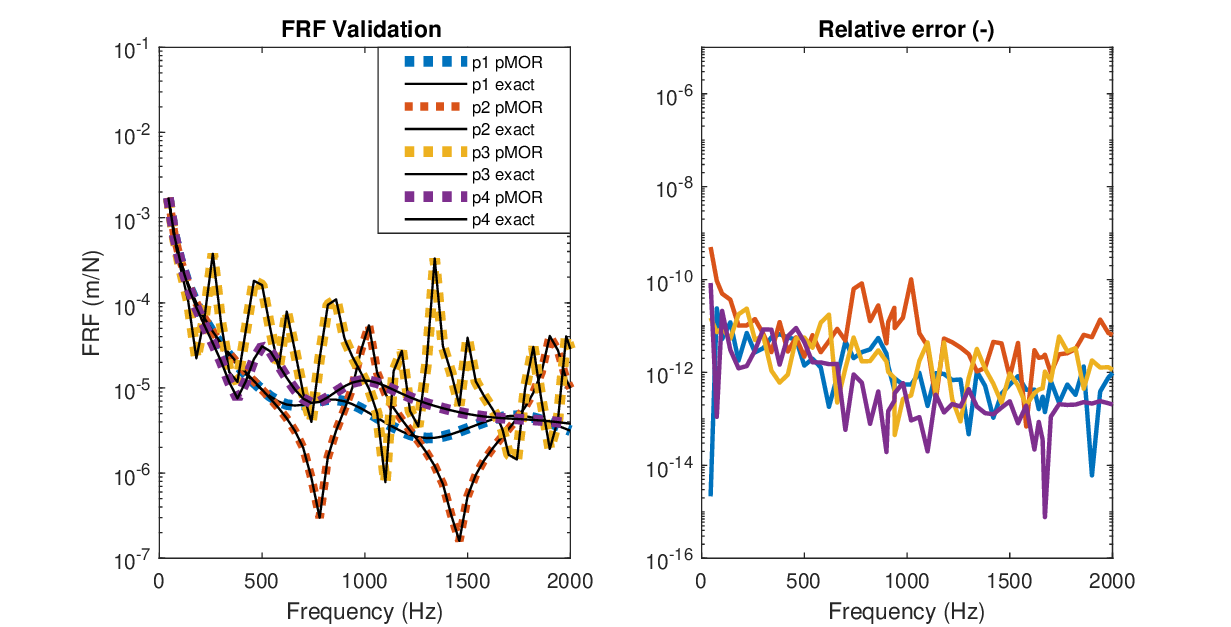}};

    \end{tikzpicture}
    \caption{FRFs of the four validation points comparing the global MOR basis with the exact viscoelastic solution (left), and the relative error between the two solutions (right).}
    \label{fig:FRF_error_Valdiationpoints}
\end{figure}

\begin{table}[h!]
\centering
\caption{Parameter bounds for the DOE.}
\begin{tabular}{|c|c|c|}
\hline
Parameter & Lower Bound & Upper Bound \\
\hline
$E_0$ (Pa) & $0.1 \times 10^9$ & $10 \times 10^9$ \\
$E_{\infty}$ (Pa) & $0.1 \times 10^9$ & $100 \times 10^9$ \\
$\tau$ (s) & $1 \times 10^{-8}$ & $0.9$ \\
$\alpha$ & $0$ & $1$ \\
\hline
\end{tabular}
\label{tab:parameter_bounds}
\end{table}
In line with the adaptive methodology proposed by~\citet{Xie2019}, we employ a quasi-random set of scrambled Sobol points to construct the reduced-order model for varying material properties~\cite{Bratley198888}. The Sobol points are described by $\mathbf{x}_{\text{sob}}$. The corresponding scaled parameter sets, $\mathbf{p}_{\text{scaled}}$, are then obtained via the transformation
\begin{equation}
    \mathbf{p}_{\text{scaled}} = \mathbf{x}_{\text{sob}} \left(\mathbf{p}_{\text{ub}} - \mathbf{p}_{\text{lb}}\right) + \mathbf{p}_{\text{lb}}.
\end{equation}
This scaling procedure ensures reproducibility and a uniform exploration of the parameter domain. In Fig.~\ref{fig:image_with_table}, a set of sampling Sobol points (blue) and validation points (red) is illustrated. Based on all these sampling design points, frequency-dependent Krylov subspaces are computed. For the structures investigated in this work 8-10 Sobol points were sufficient. Finally, the reduced-order models of all design points are truncated using singular value decomposition in the same way as shown in the previous section for the frequency-dependent base generation. The global reduced model is validated with the illustrated four design points, by comparing the reduced-order model result to the full viscoelastic calculation. The material properties of validation points are shown in the embedded table of Fig.~\ref{fig:image_with_table}. The results of the FRFs of a rectangular beam, further discussed in section~\ref{sec:POM}, are shown in Fig.~\ref{fig:FRF_error_Valdiationpoints}. A model is considered precise enough when its error is in the range of $10^{-4}$ or lower.
% , see Fig.~\ref{fig:FRF_error_Valdiationpoints}. 
The validation results show that the reduced-order model error is significantly below this benchmark. 

\section{Inverse Optimization Methodology}

\label{sec:opt}

%This study employs a constrained Particle Swarm Optimization (PSO) algorithm to optimize parameters within a reduced-order Model (ROM) framework. The key steps and components of the methodology are described below. The same parameter boundaries are applied as in the DOE explained in the previous step. 
\subsection{Cost function and optimization algorithm}
The goal of this work is to identify the viscoelastic material description by tuning its parameters until the FRF of the experiment matches the one of the MOR scheme. To identify the material properties, a loss function $g(s)$ needs to be minimized. It is defined as 
\begin{equation}
    \min_{p\in {E_0, E_{\infty},\tau,\alpha}} \sum_{i=1}^{N} g(s_i,p)=\min_{p\in {E_0, E_{\infty},\tau,\alpha}} \sum_{i=1}^{N} \Bigl( \log \bigl( \lvert h_{\text{e}}(s_i)\rvert \bigr) - \log \bigl(\lvert h_{\text{MOR}}(s_i,p)\rvert \bigr) \Bigr)^2,
    \label{eq:objectivefunc}
\end{equation}
where $h_{e}$ is the observed experimental transfer function of a measured point and $h_{MOR}$ represents the reduced-order model predictions of the numerical transfer function. The transfer functions are evaluated at discrete frequency points $s_i$. %The constant $C$ is introduced as a vertical shift factor since the excitation magnitude is not known, as described in  section~\ref{sec:exp}. 

\begin{figure}
    \centering
    \def\height{6.5cm}  % Set fixed height for images
    \def\spacing{0.05cm}  % Define spacing between images

    % Extract image widths BEFORE starting the TikZ environment

    \begin{tikzpicture}

        % First image (a)
        \node[name=plot1] at (0,0) {
            \includegraphics[height=\height]{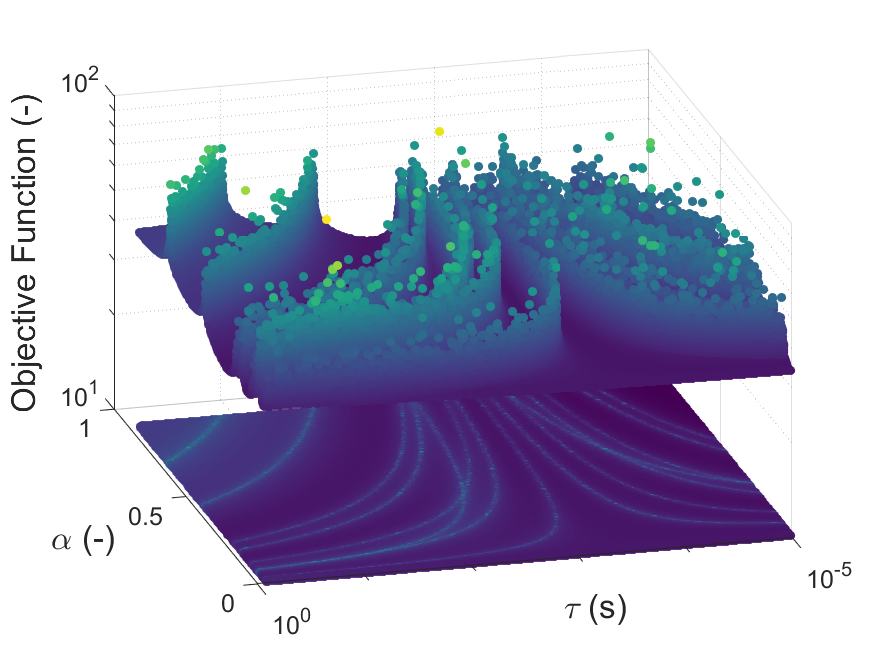}
        };
        \node[below=0.3cm of plot1] {{\centering (a) }};

        % Second image (b) next to (a)
        \node[name=plot2, anchor=west, right=\spacing of plot1] {
            \includegraphics[height=\height]{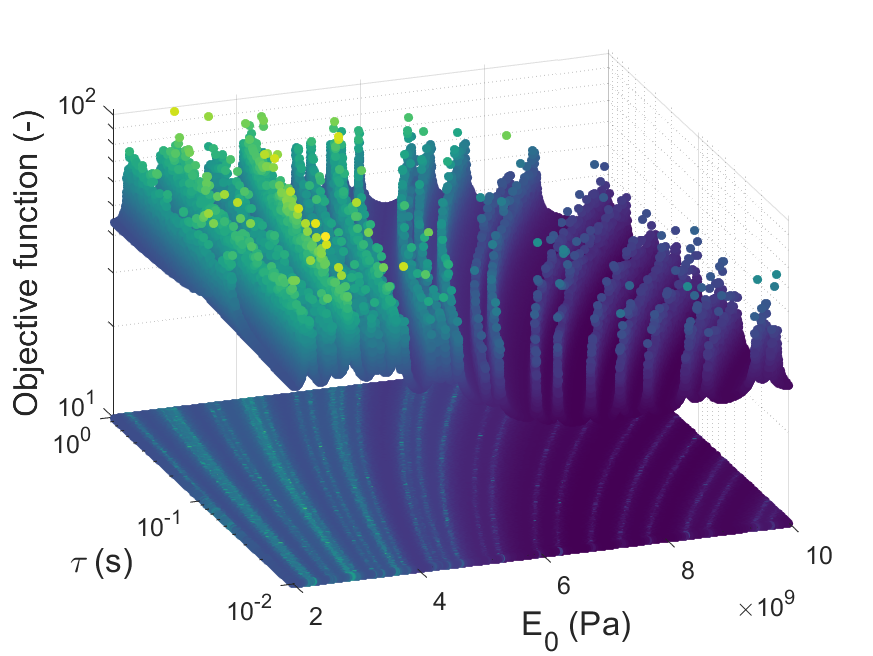}
        };
        \node[below=0.3cm of plot2] {{\centering (b) }};

    \end{tikzpicture}
    
    % Overall caption and label
 \caption{\label{fig:response_surface} Response surface visualization of the objective function (eq \ref{eq:objectivefunc}) for parameters $\alpha$ vs $\tau$ (a) and $\tau$ vs $E_0$ (b) of the polyoxymethylene beam shown in Sec.~\ref{sec:POM} using 1\,000\,000 Latin hypercube sampling points. Several local minima with similar objective function values can be distinguished.}
\end{figure}

By varying two of the four parameters, scatter plots of the loss functions are generated to illustrate the complexity of the minimization problem. Looking at a frequency range of 500-1\,000 Hz and varying e.g. the fractional order coefficient $\alpha$ and the relaxation time $\tau$, shown in  Fig.~\ref{fig:response_surface}a, or $\tau$ and the lower limit of the storage modulus $E_0$, shown in Fig.~\ref{fig:response_surface}b,  it becomes clear that gradient-based optimization schemes are not possible due to the amount of non-connected local minima. A variation of all four parameters at once would result in noisy response surfaces even if the optimization boundaries are limited to a narrow range. 

This work solves the optimization problem by using the particle swarm optimization (PSO) algorithm, implemented in MATLAB~\cite{Chen_CPSO}. PSO is selected due to its suitability for high-dimensional, nonlinear problems, and its ability to find near-optimal solutions without the need for gradient information. In addition to minimizing the objective function, the optimization can be subject to soft inequality constraints. These constraints ensure that the solution satisfies the physical limits of the viscoelastic material model. For the fractional derivative model, one obvious inequality constraint has to be fulfilled:
$E_{0} \leq E_{\infty}$.

The PSO algorithm uses a swarm of particles to explore the solution space. Each particle updates its position based on its own best solution and the best solution found by the entire swarm. The algorithm is configured with a swarm size of 200 particles. The inertia weight, which governs the balance between exploration (global search) and exploitation (local search), is set to 0.9. We use a cognitive coefficient (\(c_1=0.5\)) and social coefficient (\(c_2=1.25\)). These influence how much each particle relies on its own experience and the swarm's experience. Particle velocities are clamped between [-1, 1] to avoid large jumps in the solution space.

The stopping criterion is defined based on the improvement in the global best solution: the algorithm terminates when the average relative change over all solutions is smaller than \( 10^{-3} \) over 5 consecutive iterations, or after reaching the maximum of 100 iterations. 
The computational effort caused by swarm size and iterations leads in this specific case to 20\,000 computations of FRFs. Since the calculation time of a frequency response function using the full model as posed in eq.~\ref{eq:FRF} lies in the range of minutes, such an optimization problem with several thousands of function calls can only be solved with the MOR approach. After calculating the MOR base, a single frequency response calculation only takes several milliseconds.

\subsection{Experimental setup for the acquisition of  FRFs}
\label{sec:exp}
Since the loss function $g(s)$ requires experimental data, the measurement setup needs to be accurate and reproducible. Vibrometric experiments are conducted using an Optomet SWIR SLDV. A surface grid is constructed and in each measurement point the vibrational velocity is measured. The excitation is achieved via an electromagnet of type ITS-MSM-1212-24VDC with a maximum voltage of 24 Volt, power of 1.4 Watt and retention force of 20~N. The magnet has a diameter of 15 mm. The electromagnet acts on a small permanent magnet glued to the sample surface. This cylindrical magnet has a weight of 0.2 g, a diameter of 5 mm and a height of 1.5 mm. This adds minimal weight to all the samples used in this study. For all measurements we excite the samples with a sine sweep ranging from 10 to 10\,000 Hz. The boundary conditions are chosen to be free, since this yields the best agreement with numerical simulations, and the entire setup is mounted on a vibration isolation table. Both the excitation and the measurement are non-contact methods, thereby avoiding local changes in mass and stiffness that might alter the inverse optimization problem.

\begin{figure}
    \centering
    \includegraphics[trim=5 8 5 18, clip,width=0.9\linewidth]{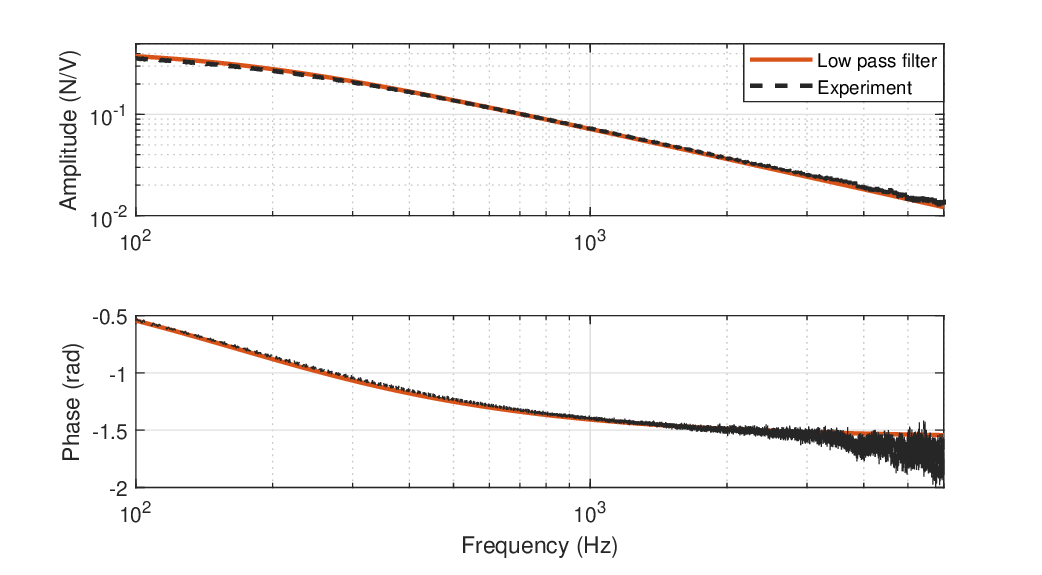}
    \caption{Measured voltage to force response function (black dashed lines) and theoretical (normalized) approximation by a low pass filter with a cut-off frequency of 1040~Hz (orange line).}
    \label{fig:lowpass}
\end{figure}

Since the magnetic field strength is not expected to remain constant over the frequency range, an additional measurement was conducted with the permanent magnet mounted on a force sensor. This revealed a linear low pass behavior of the electromagnet, with an angular cut-off frequency $\omega_0=1040$~Hz, as shown in Fig.~\ref{fig:lowpass}. The measured voltage to force response function is required to calculate the numerical FRF since the model assumes a constant force excitation.

\section{Demonstration of the inverse material identification}
In this section, two different use cases are presented. First, beams made of polyoxymethylene (POM) are identified  using the inverse strategy presented in the previous sections, and the results are validated by standard DMA measurements. The identification process allows to compare three samples from different manufacturers to results from accurately machined samples that fit in the climate chamber of the DMA machine. As a second application, a more complex curved sample produced through additive manufacturing is examined, composed of a polymer–ceramic mixture. Additively manufactured materials are not suited for standard DMA measurements, due to the need of machined specimens, showcasing the advantage of our proposed methodology. 
Both finite element models exhibit comparable DOF, approximately in the range of 12\,000–15\,000 for the POM sample. For a computational comparison between FE model and MOR, the forced vibration problem is solved 400 discrete frequency steps for selected frequency ranges between 50–6\,000 Hz, ensuring sufficient resolution for harmonic response analysis. In a non-parallelized MATLAB environment running on an Intel i7 processor (3.6 GHz) with 32GB RAM, the full finite element solution requires between 500–650 seconds per simulation, illustrated in Tab.~\ref{tab:time_measuremetns}.
For comparable computations, a reduced-order model using 150 basis vectors was implemented, which has proven to be enough for material identification. This basis size decreases the computation time to approximately 0.05 seconds per simulation, enabling up to 10\,000 evaluations across varying material parameters in the same time as one FE calculation. Such efficiency is particularly advantageous for optimization tasks that involve iterative fine-tuning of material models.

\begin{table}[h]
\caption{Computational overview and optimization variables for the two investigated use cases.}
\centering
\setlength{\tabcolsep}{10pt}
\renewcommand{\arraystretch}{1.2}
\begin{tabular}{l r r r r}
\toprule
 & \multicolumn{1}{c}{FE} &
 \multicolumn{1}{c}{MOR} & \multicolumn{1}{c}{population size} &
 \multicolumn{1}{c}{iterations} \\
 & \multicolumn{1}{c}{\small 200 freq. steps}  &
 \multicolumn{1}{c}{\small 200 freq. steps} &
 \multicolumn{1}{c}{\small } & \\ 
\midrule
POM         & 650  & 0.05 & 200 & 100 \\
3D-Print  & 500 & 0.05 &  200 &  70 \\
\bottomrule
\end{tabular}
\label{tab:time_measuremetns}
\end{table}

\begin{figure}[h!]
    \centering
    \begin{tikzpicture}
        % First image (POM samples)
        \begin{axis}[
            name=plot1,
            hide axis,  % Hide the axis entirely
            height=5cm, width=8cm,  % Adjust these as needed
            xmin=0, xmax=1749, ymin=0, ymax=1749  % Image dimensions
        ]
        \addplot [forget plot] graphics 
            [xmin=0, xmax=1749, ymin=0, ymax=1749]
            {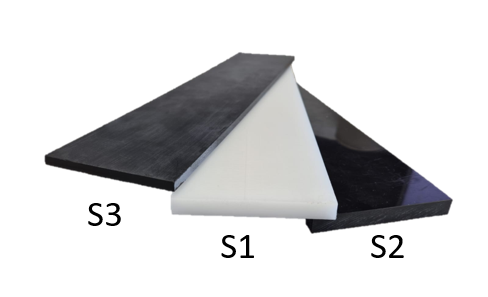};
        \end{axis}
        \node[below=0.3cm of plot1] {(a)};

        % Second image (3-point bending beams), positioned 1 cm to the right
        \begin{axis}[
            name=plot2,
            at={($(plot1.east)+(0.3cm,0)$)},  % Position 1cm to the right of plot1
            anchor=west,  % Align the west edge
            hide axis,  % Hide the axis
            height=5cm, width=4cm,  % Adjust these as needed
            xmin=0, xmax=1749, ymin=0, ymax=1749
        ]
        \addplot [forget plot] graphics 
            [xmin=0, xmax=1749, ymin=0, ymax=1749]
            {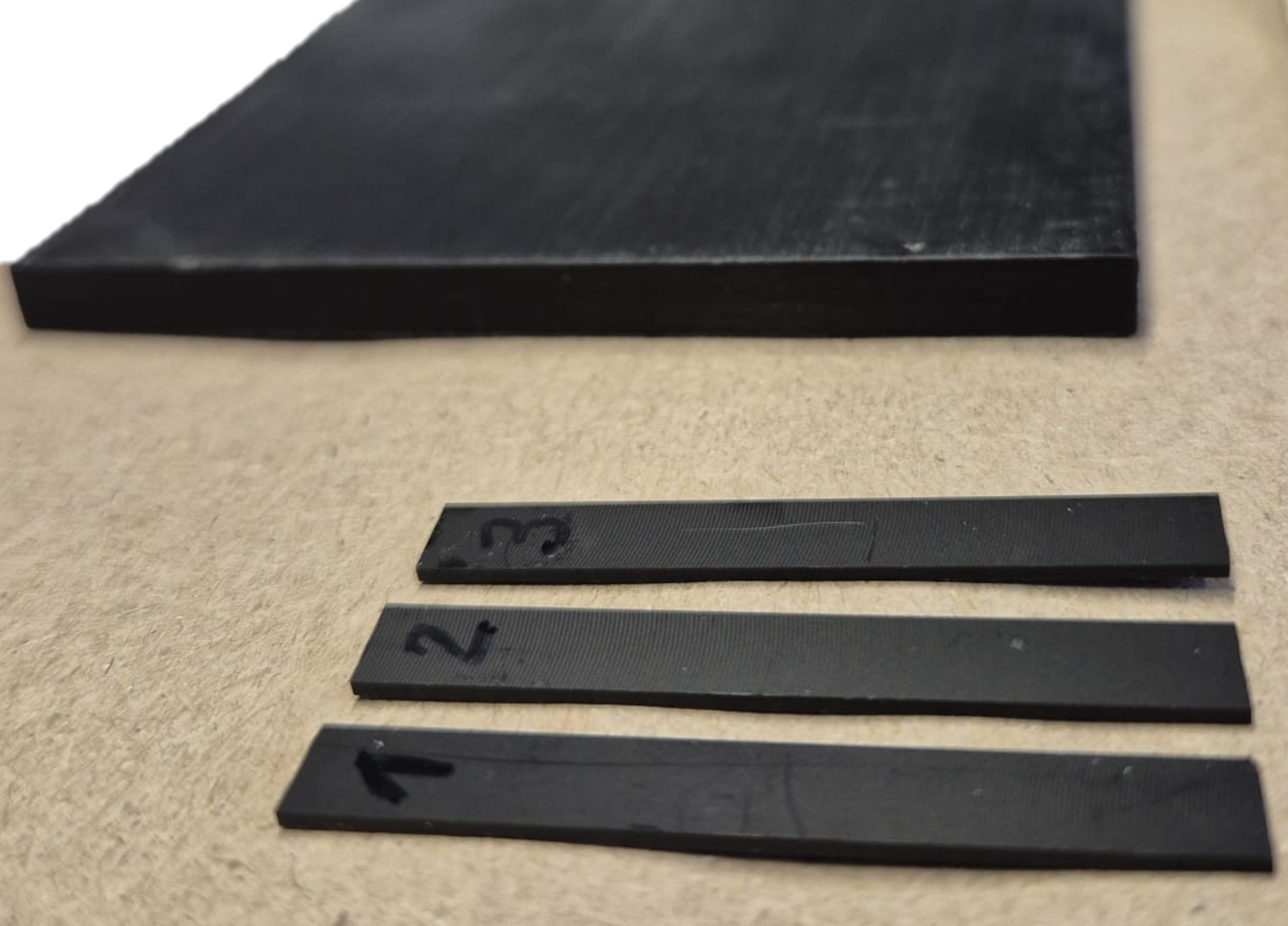};
        \end{axis}
        \node[below=0.3cm of plot2] {(b) };

         \begin{axis}[
            name=plot3,
            at={($(plot2.east)+(0.7cm,0)$)},  % Position 1cm to the right of plot1
            anchor=west,  % Align the west edge
            hide axis,  % Hide the axis
            height=5cm, width=8cm,  % Adjust these as needed
            xmin=0, xmax=1749, ymin=0, ymax=1749
        ]
        \addplot [forget plot] graphics 
            [xmin=0, xmax=1749, ymin=0, ymax=1749]
            {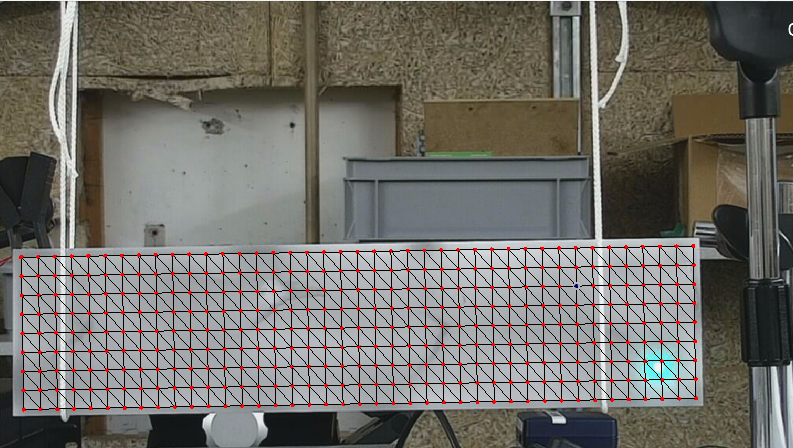};
        \end{axis}
        \node[below=0.3cm of plot3] {(c) };

      % Cross node using cross out
   
    \node[draw=blue, circle,fill=blue, line width=0.05pt,inner sep=0pt, minimum size=0.1cm, label={[draw=black, fill=white, inner sep=1pt, font=\tiny, scale=0.99, text=blue]above left:{ID=1}}] at (10.01,1.46) {};
    
      % Empty node (visible as a circle)
    \node[draw=blue, circle, fill=blue, line width=0.05pt, inner sep=0pt,minimum size=0.1cm,label={[draw=black, fill=white, inner sep=1pt, font=\tiny, scale=0.99, text=blue]above left:{ID=2}}] at (10.01,0.3) {};
        
    \node[cross out, draw=white, line width=0.5pt, minimum size=0.1cm,inner sep=0pt] at (10.01,1.46) {};
    \end{tikzpicture}
    
    \caption{Three POM samples from different manufacturers  (a), 3-point bending DMA test specimens (b) and experimental setup (c): Measuring grid of the SLDV off the investigated structure which is measured approximating free boundary conditions by hanging beam configuration. The white cross indicates the excitation position on the back side and the two blue points (ID 1 and 2) are analyzed for material identification.}
    \label{fig:POM_samples}
\end{figure}
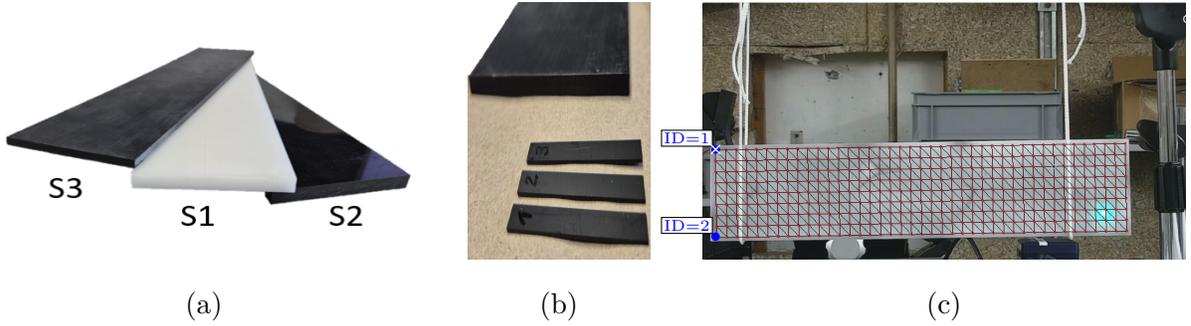

\subsection{Usecase I: Polyoxymethylene}
\label{sec:POM}

The investigated POM samples shown in Fig.~\ref{fig:POM_samples}a are of dimensions $350 \times 100 \times 10$~mm$^3$ and have a density of $1.41 \,\text{g/cm}^3$. The numerical model comprises approximately 12\,000 DOF using shell elements, for which the mesh convergence is validated up to 6\,000 Hz. To approximate free boundary conditions the beams are hung up using soft strings, shown in Fig.~\ref{fig:POM_samples}c. The laser is aligned to enable the measurement of normal surface velocities. The sample is excited on the top left corner on the back side. The material properties are retrieved by standard DMA measurements and by the proposed inverse identification.

\begin{figure}[h!]
\begin{tikzpicture}
\centering
\def\width{8cm};  % Define a fixed width for both images
\def\spacing{0cm}; % Define spacing between images
\def\height{7cm};

    % First image (DCM)
    \node[name=plot1] at (0,0) {\includegraphics[height=\height,trim=20 0 80 0, clip]{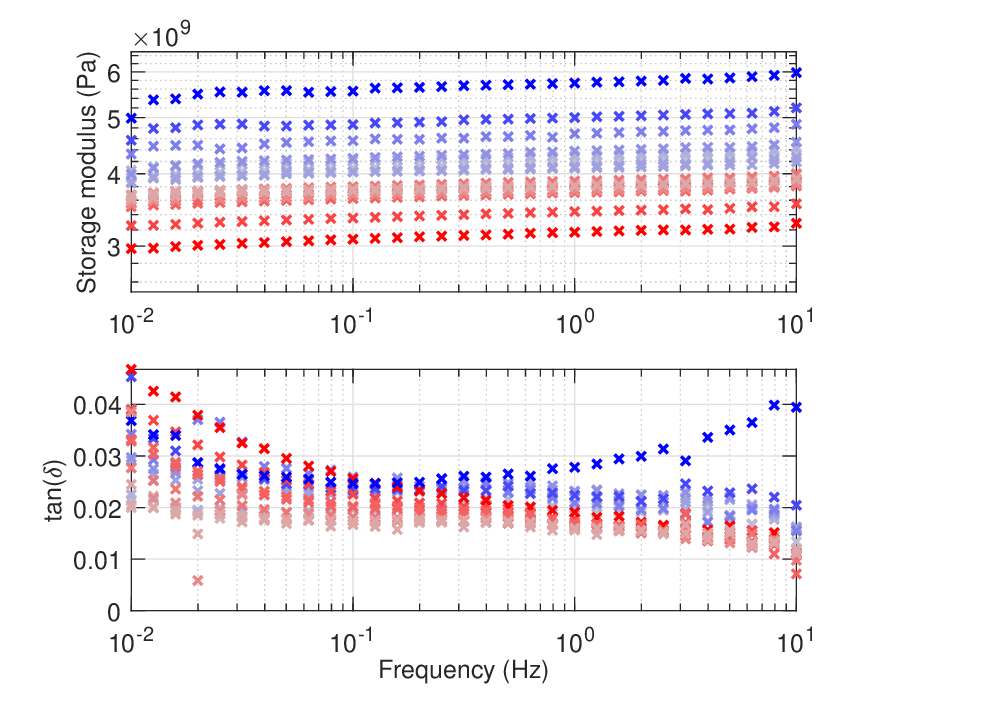}};

        %{\includegraphics[height=\height,trim=0 0 70 0, clip]{fig/fig_DMA/data/DMA_raw.eps}};

    \node[below=-0.1cm of plot1] {{\centering (a) }};

    % Second image (BEM) next to the first one
    \node[anchor=west, right=\spacing of plot1] (plot2) 
        {\includegraphics[height=\height,trim=20 0 20 0, clip]{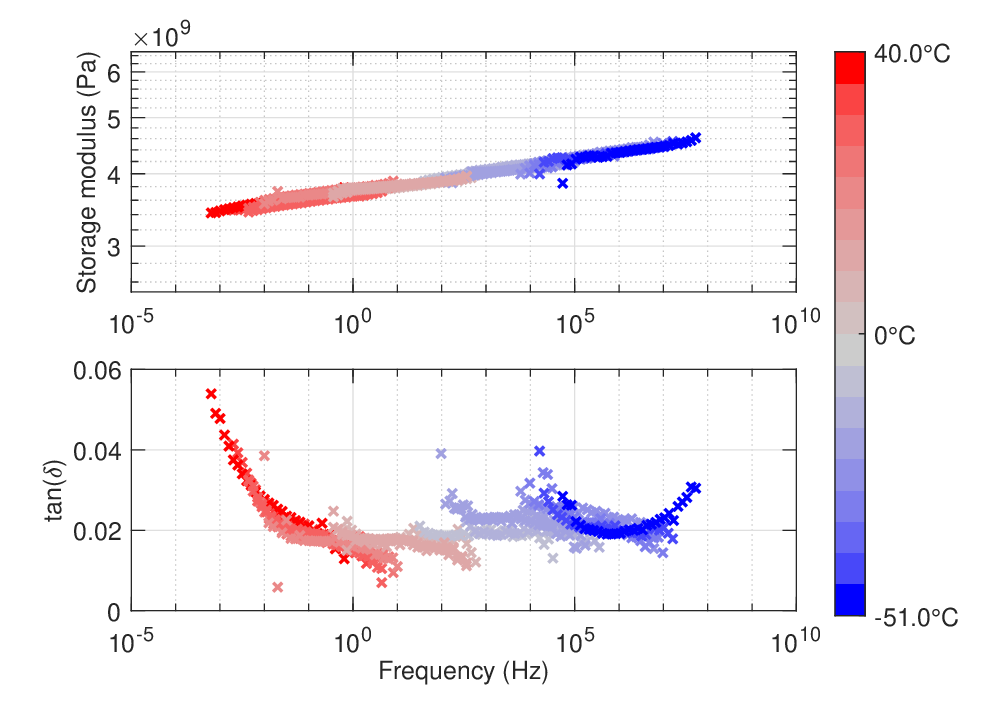}};
        %{\includegraphics[height=\height]{fig/fig_DMA/data/DMA_shift.eps}};

    \node[below=-0.1cm of plot2] {{\centering (b) }};

\end{tikzpicture}
\caption{Raw  data of DMA measurements at different temperatures (left) and the shifted data in frequency after applying time-temperature superposition (right).}
\label{fig:DMA}
\end{figure}

\subsubsection{Dynamical mechanical analysis}
Viscoelastic materials are commonly characterized using DMA, which can typically perform measurements within a frequency range of 0.01 to 100 Hz. However, this does not capture the full frequency-dependent material complexity for the range in which the material is operated—especially at the extreme low or high ends. Therefore, time-temperature superposition (TTS) has been used to extend the covered frequency range. In this work, all measurements are performed using a TSA III Theometric System Analyzer (TA Instruments, USA), ensuring precise control of temperature and measurement parameters.

In order to create a benchmark for a single POM sample, small laser-cut samples of size $45 \times 6 \times 1$~mm$^3$ (shown in Fig.~\ref{fig:POM_samples}b) are excited by a strain sweep doing three-point bending tests at temperatures from -51°C to 40°C in the frequency range from 0.01-10 Hz since at higher frequencies the simply supported boundary conditions become unstable in the case of such a stiff material.  In Fig.~\ref{fig:DMA}a, it is shown how the storage modulus increases approximately linearly with frequency,  shifting upward at lower and downward at higher temperatures. These data can then be shifted in the frequency domain using the time-temperature superposition method, where the data is shifted in magnitude and frequency~\cite{TAOrchestrator2008}. As a reference temperature, the measurement curve at 20°C is selected. Since the measured phase is very small ($\tan(\delta) < 3\%$), the phase shift measurement in the time domain is more prone to errors due to the low signal-to-noise ratio.
%The low damping indicated by the low phase, also makes the probes more affected to unstable boundary conditions. 
Thus, the master curve considering the overall frequency range is not smooth, as can be seen in Fig.~\ref{fig:DMA}b. 

% First plot (DCM)
\begin{figure}[h!]
\centering
\begin{tikzpicture}
  \def\imgwidth{16cm}
  \node[anchor=south west,inner sep=0] at (0,0) {%
    \includegraphics[width=\imgwidth]{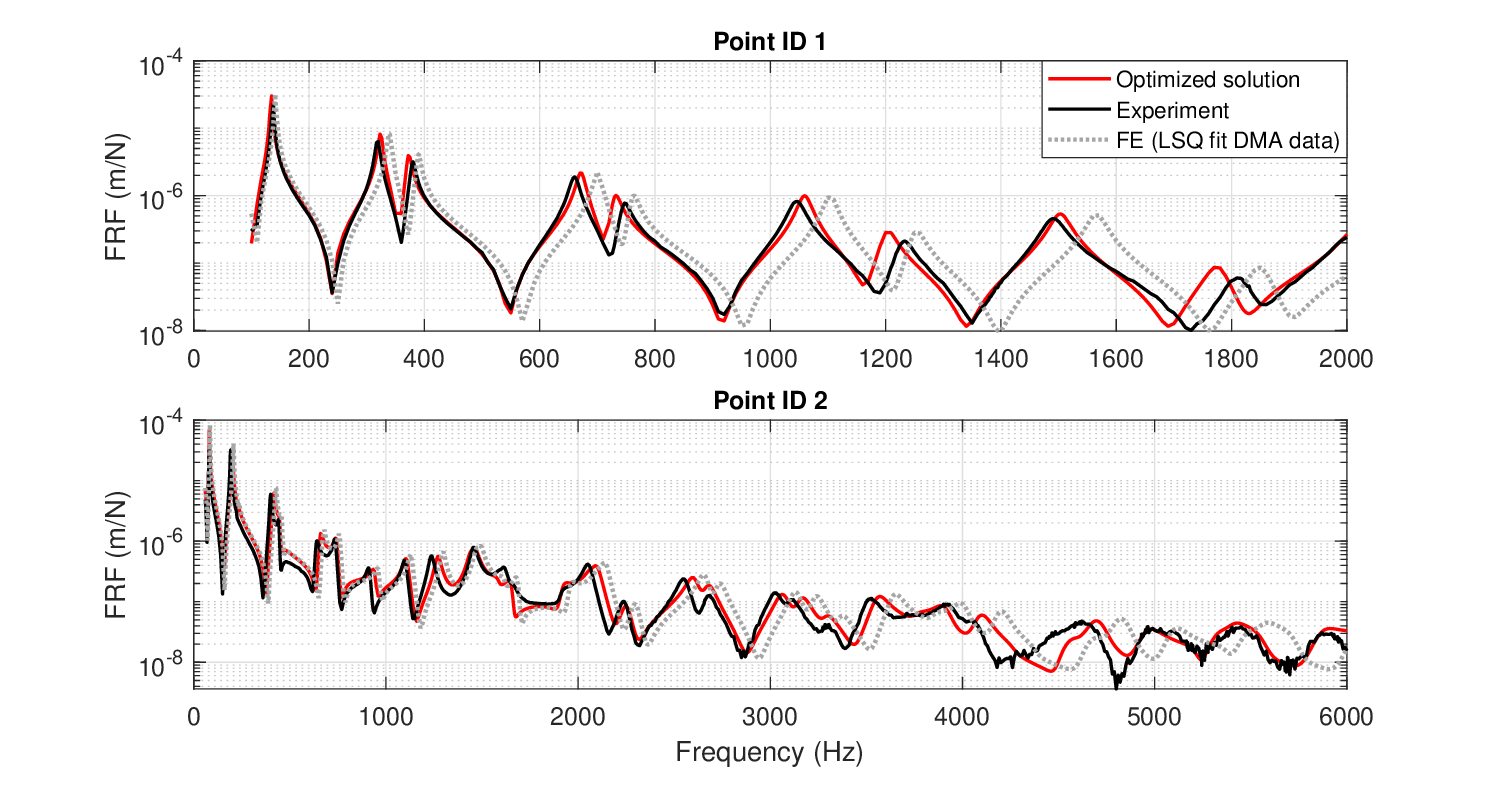}};
\end{tikzpicture}
\caption{\label{fig:fit_frf_ID1} Measured and simulated FRF using the optimized viscoelastic material model for a frequency range of 100-2\,000 Hz (upper) for node 1 and for a frequency range of 100-6\,000 Hz (lower) for node 2. Additionally the FE analysis using DMA data is performed, illustrating how much the results deviate.}
\end{figure}

\begin{figure}[h!]
\centering
\begin{tikzpicture}
  \def\imgwidth{14cm}
  \node[anchor=south west,inner sep=0] at (0,0) {%
    \includegraphics[width=\imgwidth,trim=0 20 0 20, clip]{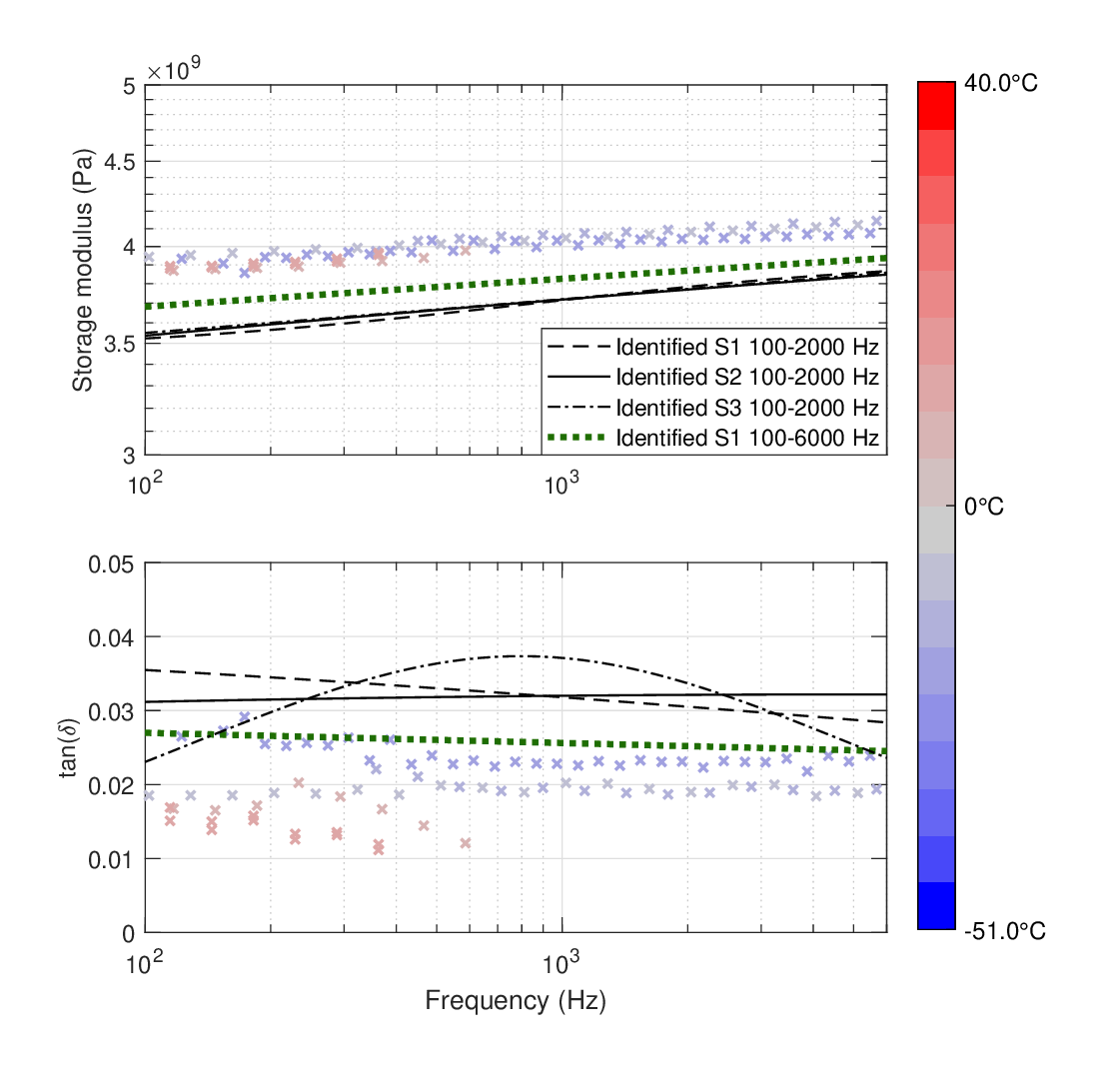}};
\end{tikzpicture}
\caption{Comparison of the shifted DMA measurements at different temperatures to three identified material models of different POM-beams of different manufacturers for the frequency range of 100-2\,000 Hz (Point 1). Additionally also the optimized solution for Point 2 up to 6\,000 Hz is shown. }
\label{fig:DMA_compareID1}
\end{figure}

\begin{table}[h!]
\centering
\caption{Comparison of resonance frequencies and FWHM.}\label{tab:peakComparison}
\begin{tabular}{c c c c c c}
\hline
$f_{exp}$ & $f_{MOR}$& f rel. error (\%) & Exp. FWHM & 
MOR FWHM & FWHM rel. error (\%) \\ \hline
\begin{comment}
138.0 & 135.0 & 2.2 & 6.2 & 6.0 & 3.3 \\ 
320.0 & 323.0 & 1.0 & 14.2 & 12.5 & 12.2 \\ 
380.0 & 372.1 & 2.1 & 15.8 & 15.3 & 3.5 \\ 
660.0 & 670.1 & 1.5 & 27.9 & 26.1 & 6.4 \\ 
747.1 & 732.1 & 2.0 & 32.7 & 29.9 & 8.7 \\ 
1047.2 & 1059.2 & 1.1 & 47.3 & 39.1 & 17.2 \\ 
1233.2 & 1200.0 & 2.7 & 49.4 & 48.3 & 2.2 \\ 
1490.0 & 1504.2 & 1.0 & 63.6 & 55.1 & 13.4 \\ 
1809.3 & 1770.0 & 2.2 & 63.1 & 58.3 & 7.6 \\ 
\end{comment}
138.0 & 135.0 & 2.2 & 0.045 & 0.044 & 3.97 \\ 
320.0 & 323.0 & 1.0 & 0.044 & 0.039 & 12.24 \\ 
380.0 & 372.1 & 2.1 & 0.042 & 0.041 & 4.08 \\ 
660.0 & 670.1 & 1.5 & 0.042 & 0.039 & 6.57 \\ 
747.1 & 732.1 & 2.0 & 0.044 & 0.041 & 8.93 \\ 
1047.2 & 1059.2 & 1.1 & 0.045 & 0.037 & 17.23 \\ 
1233.2 & 1200.0 & 2.7 & 0.040 & 0.040 & 3.48 \\ 
1490.0 & 1504.2 & 1.0 & 0.043 & 0.036 & 13.44 \\ 
1809.3 & 1770.0 & 2.2 & 0.035 & 0.033 & 7.91 \\ 
\hline
\end{tabular}
\end{table}

\subsubsection{Inverse identification applied to vibrometric experiments}

Next, we perform the inverse material identification for the POM sample. The transfer functions of two points, shown in Fig.~\ref{fig:fit_frf_ID1}, are fitted. For the location near the excitation position (ID 1) a frequency range of 100-2\,000 Hz with 400 linearly spaced frequency steps was investigated. The second point (ID 2) was optimized for a larger frequency range of 50-6\,000 Hz with 800 frequency steps. This point was selected due to broad peaks at higher frequencies. In the Appendix, details on the sensitivity of force and receiver location are further explained.  Several features of the measured and experimental FRF can be used to quantify the accuracy of the material model, as shown in Tab.~\ref{tab:peakComparison}. Peak frequency measurements, which correlate with the material stiffness, agree within an error margin of less than 3\% between optimized solution and experiment. However, the torsional mode shapes, e.g. near 1\,200 Hz, exhibit less accurate matching compared to the bending modes, possibly due to the non-perfect string boundary conditions. As shown in the Appendix, the torsional vibration peaks are also very sensitive to the location of the excitation force, so that larger deviations can be expected. Peak widths are determined using the full width at half maximum (FWHM) values computed by the \textit{findpeaks} function in MATLAB and normalized with respect to the peak frequency, show errors ranging from 3\% to 17\%. The higher relative errors are found for lower peaks, where the FWHM method is likely to result in less accurate values because of closely spaced resonances and the resolution of the frequency response. 

The optimized material model is compared with the DMA measurements in Fig.~\ref{fig:DMA_compareID1}. The results indicate a discrepancy of approximately $10\%$ between the DMA measurements and the identified Young's modulus, which may be attributed to the heat treatment of the small samples during laser cutting, or to inaccuracies in the DMA bending experiments. Further error is introduced due to the fact that phase measurements are close to the tolerance limit. Both are a potential problem in using the DMA data for the prediction of the dynamic response of larger structures. This is illustrated by the FRFs calculated with DMA material data shown in Fig.~\ref{fig:fit_frf_ID1}: the dotted line is clearly shifted with respect to the measured FRF. By incorporating a least square (LSQ) fit of the the DMA data into the numerical model, a shift in frequency of approximately 10\% can be observed.  Nevertheless, the agreement in the slopes of both the storage modulus and loss factor demonstrates that the identification method is a promising approach to determine frequency-dependent properties without the need for DMA sample preparation. 

In a second setup, a point (ID 2) with less pronounced anti-resonance behavior is optimized for frequency range from 50-6\,000 Hz, as shown in Fig.~\ref{fig:fit_frf_ID1}. The agreement at high frequencies appears even better from a visual perspective, as broader peaks can be considered, allowing to include more structural characteristics of the investigated system.
%To assess the accuracy of the identified material model, a full finite element model is compared to the reduced-order model using the global basis. The results demonstrate that the relative error remains below $10^{-5}$  across the entire frequency range, thereby confirming the robustness of this method for a wide range of material models and frequency spectra. 
Fig.~\ref{fig:DMA_compareID1} shows that the storage modulus deviation from the wider frequency FRF differs approximately 5$\%$ for low frequencies and 3$~\%$ for high frequencies from the DMA results. A comparison between the two identified frequency ranges—narrowband (100–2\,000 Hz) and broadband (50–6\,000 Hz)—reveals slight deviations, highlighting the importance of optimizing the relevant frequency range to ensure reliable data. For the identification of a detailed master curve, a wide spectrum is needed, at the cost of localized offsets. There for the selected frequency band is strongly dependent on the application. Additionally, it becomes evident, that the fractional derivative model is not detailed enough with its four parameters. %The damping coefficient ranges from 0.024 to 0.027 and therefore agrees well with the temperature measurement at -10$^\circ$C. 

\subsection{Usecase II: 3D-printed curved samples made of composite material}
\label{sec:CS}

\begin{figure}[t]
  \centering
  \begin{tikzpicture}[node distance=0cm and 0cm]
    %────────────────────────────────────────────────────────────────
    % Panel (a): Granular structure + scale bar
    %────────────────────────────────────────────────────────────────
    \node (plot1) {%
      \begin{tikzpicture}[inner sep=0, outer sep=0]
        % the image
        \node[anchor=south west] (image) at (0,0)
          {\includegraphics[
             width=0.31\linewidth,
             trim=80 0 0 0,
             clip
           ]{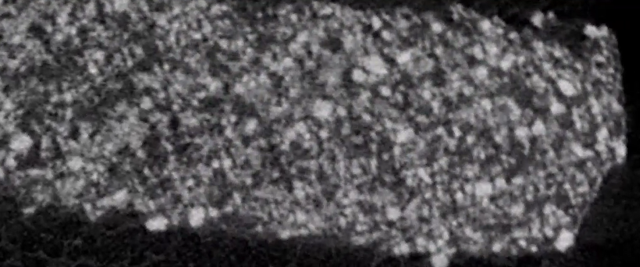}};
        % overlay the scale bar relative to image.south west
        \begin{scope}[shift={(image.south west)}]
          \begin{scope}[xshift=0.5cm, yshift=0.5cm]
            \draw[red, ultra thick] (0,0) -- ++(1,0);        % 4 cm bar
            \draw[red, ultra thick] (0,-0.25) -- (0,0.25);      % left tick
            \draw[red, ultra thick] (1,-0.25) -- (1,0.25);      % right tick
            \node[red] at (0.5,0.75) {\LARGE\textbf{2mm}};     % label
          \end{scope}
        \end{scope}
      \end{tikzpicture}%
    };

    %────────────────────────────────────────────────────────────────
    % Panel (b): Experimental setup
    %────────────────────────────────────────────────────────────────
    \node (plot2) [right=0cm of plot1] {%
      \includegraphics[
        width=0.3\linewidth,
        trim=50 0 0 0,
        clip
      ]{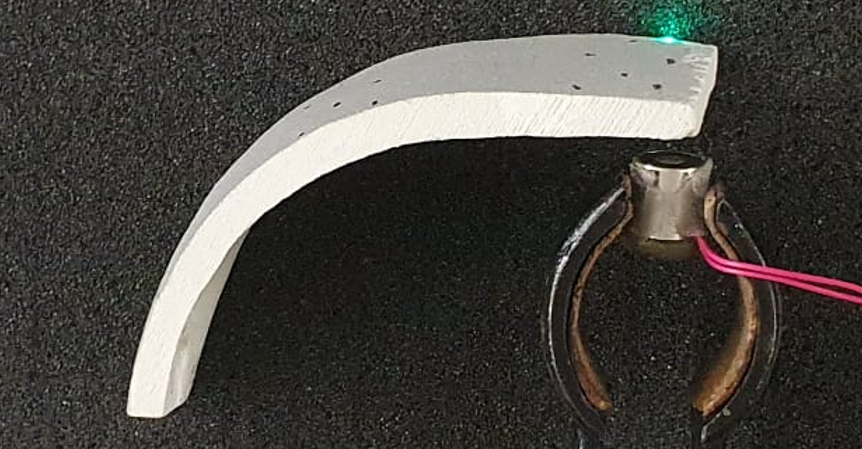}%
    };

    %────────────────────────────────────────────────────────────────
    % Panel (c): FE model
    %────────────────────────────────────────────────────────────────
    \node (plot3) [right=0cm of plot2] {%
      \includegraphics[
        width=0.3\linewidth,
        trim=0 0 0 0,
        clip
      ]{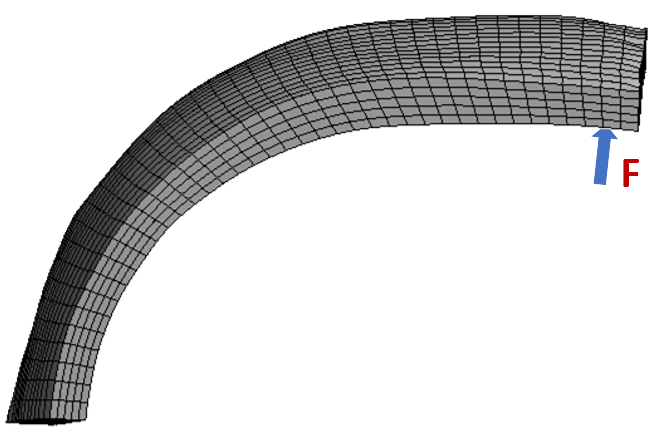}%
    };

    %────────────────────────────────────────────────────────────────
    % Subfigure labels
    %────────────────────────────────────────────────────────────────
    \node[above=-2.0mm of plot1] {(a)};
    \node[above=-2.0mm of plot2] {(b)};
    \node[above=-3.7mm of plot3] {(c)};
    % Empty node (visible as a circle)
      \node[draw=blue, circle, fill=blue, line width=0.05pt, inner sep=0pt,minimum size=0.1cm] at (13.44,1.2) {};
      \node[draw=blue, circle, fill=blue, line width=0.05pt, inner sep=0pt,minimum size=0.1cm] at (6.88,0.95) {};
  \end{tikzpicture}
\vspace{-0.5cm}
  \caption{%
    (a) Granularity in CT scan of polymer ceramic composite structure.
    (b) Experimental setup of the excited structure with an electromagnet.
    (c) Corresponding numerical (FE) model. The blue dots indicate the measurement position of the displacement in Laser beam direction. 
  }
  \label{fig:PhaconPatch}
\end{figure}
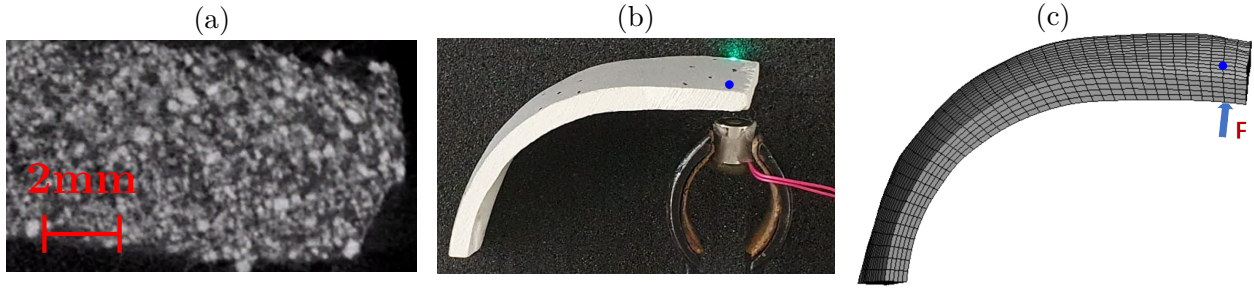

The homogeneous material analyzed in the previous section allows a comparison with standard DMA measurements. In addition we examine a 3D-printed composite material commonly used to replicate human skulls. These replicas are employed by surgeons-in-training to test various procedures, as the material mimics bone properties. The composite consists of a polymer-ceramic blend and, while it appears structurally heterogeneous at the microscale -- as observed in CT scans (Fig.~\ref{fig:PhaconPatch}a) -- numerical models require a homogenized material description. The material distribution is influenced by the additive manufacturing process known as Powder Bed Binder Jetting, which potentially changes the macroscopic homogenized density and stiffness properties. Therefore, costly and time-consuming DMA tests are not suitable to be carried out on every new sample whereas dynamic testing of various printed samples yields a fast and non-destructive solution to determine the material properties.

\begin{figure}[h!]
    \centering
  
    \includegraphics[width=0.42\linewidth,trim=280 220 270 60 ,clip]{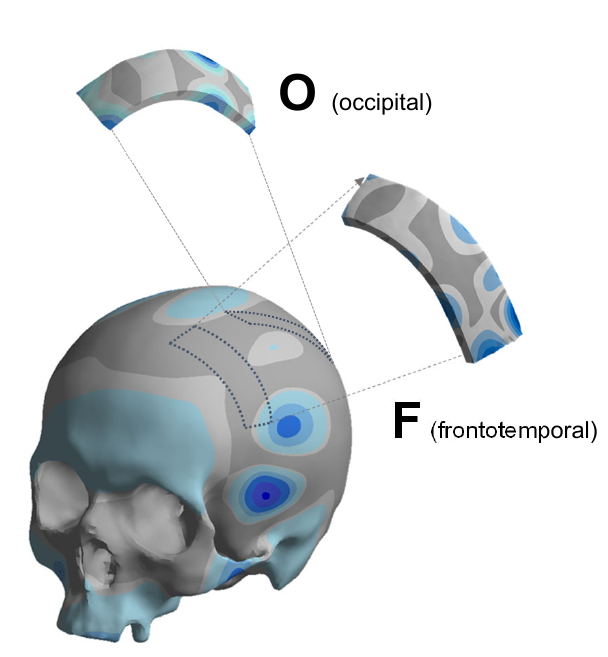}
    \hfill \includegraphics[width=0.57\linewidth,trim=0 0 0 0 ,clip]{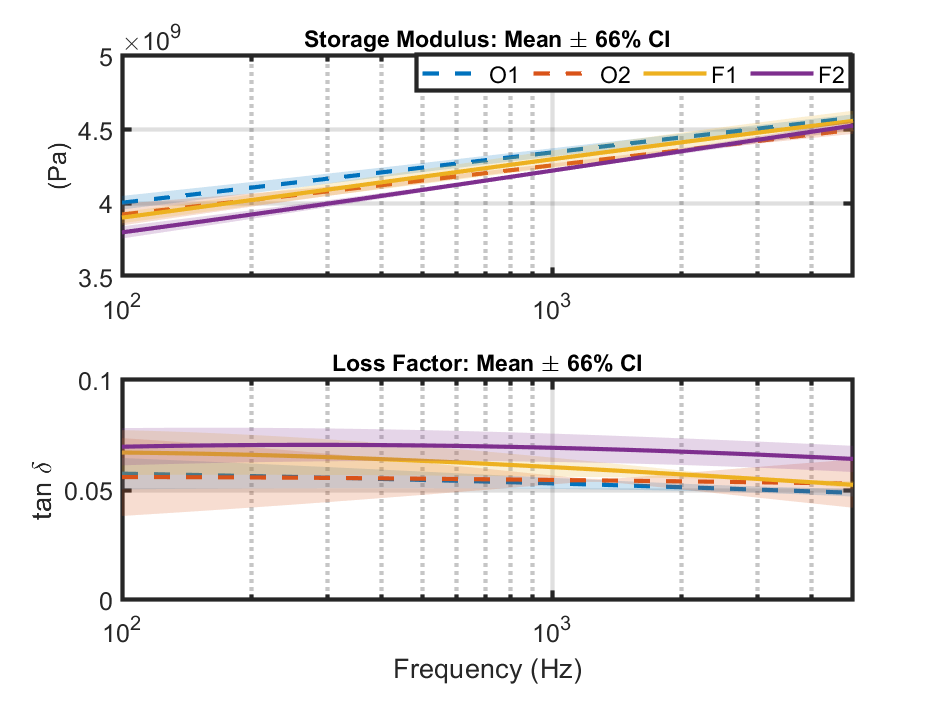}%{images/IdentifiedViscoelastic_models_Phacon5.eps}
    
    \caption{Bone patch geometries of a 3D printed head. The samples are named by their location, \textbf{F} for frontotemporal bone and \textbf{O} for occipital bone. }
    \label{fig:Phacon_head_patch}
\end{figure}

For the purpose of this study we analyze 3D-printed bone patches extracted from various locations on the skull, as shown in Fig.~\ref{fig:Phacon_head_patch}, and characterize the material properties of these regions. The patches have lengths ranging from 12 to 14 cm, widths between 3 and 4 cm, and thicknesses varying from 1 to 10 mm. The experimental and simulation setups are presented in Fig.~\ref{fig:PhaconPatch}b and Fig.~\ref{fig:PhaconPatch}c, respectively.

In the experimental setup, each specimen is supported by a  lightweight foam to approximate free boundary conditions.  The numerical model comprises approximately 15\,000 DOF using linear elements, for which the mesh convergence is validated up to 10\,000 Hz. Sample masses are listed in Tab.~\ref{tab:sample_weights}, confirming that the magnet contributes less than 1\% to the total mass.
\begin{figure}[htbp]
  \centering
\begin{tikzpicture}
\def\height{7.3cm};
\def\width{10cm};
    % First image (DCM)
    \begin{axis}[name=plot1,
        hide axis,  % Hides the axis entirely (no labels, ticks, etc.)
        height=\height, width=1*\width,  % Set the desired height and width
        xmin=0, xmax=1749, ymin=0, ymax=1749  % Image dimensions
    ]
    \addplot [forget plot] graphics [xmin=0,xmax=1749,ymin=0,ymax=1749]{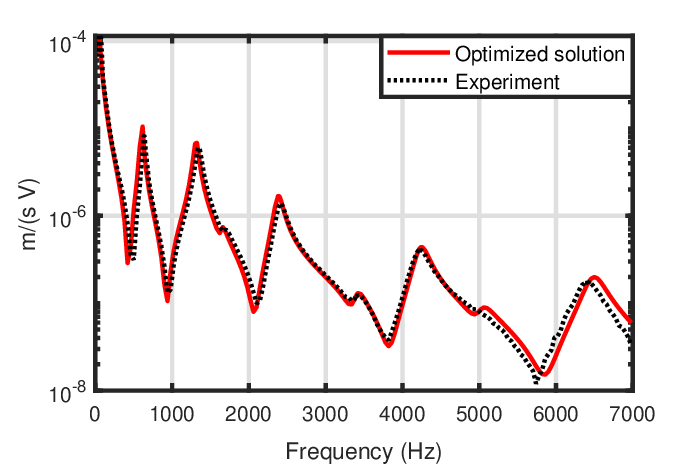};
    \end{axis}

    % Second image (BEM) next to the first one
    \begin{axis}[name=plot2,
        at={($(plot1.east)+(0cm,0)$)},  % Position 1cm to the right of the first plot
        anchor=west,  % Align the west edge
        hide axis,  % Hides the axis entirely
        height=\height, width=1*\width,  % Set the same height and width
        xmin=0, xmax=1749, ymin=0, ymax=1749  % Image dimensions
    ]
    \addplot [forget plot] graphics [xmin=0, xmax=1749, ymin=0, ymax=1749] {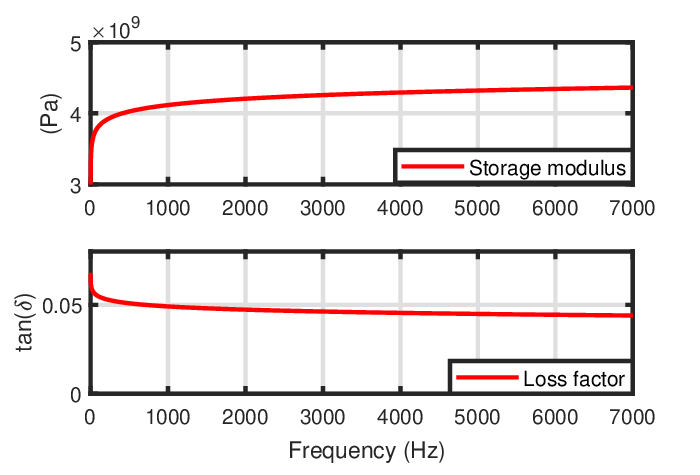};
    \end{axis}

\end{tikzpicture}
\caption{\label{fig:CS_identification} Frequency response with optimized material model compared to experiments (left). Storage modulus and loss factor of the optimized material model (right).}
\end{figure}

To illustrate the results, we present the FRF of a point near the excitation location and the identified material model for sample F1, as shown in Fig.~\ref{fig:CS_identification}. Our analysis focuses on the audible frequency range, specifically from 100 to 7\,000 Hz. Within this range, eight distinct resonances were identified, and the model fits the data with frequency and damping errors below 1\%. Notably, even the anti-resonances -- typically difficult to replicate -- show excellent agreement. Similar to the previous case, pronounced resonances at lower frequencies are captured with slightly lower accuracy. %For the full optimization cycle 10,000 function calls were performed.

To further test the robustness of the method, we analyzed four skull patches, two of each location. For each sample, measurements were taken at five points near the excitation location (black dots in Fig.~\ref{fig:PhaconPatch}b). A material model was identified for each of the five points, and the average and standard deviation were computed. This process was repeated for all four samples.
\begin{table}[h!]
  \centering
    \caption{Weights of the investigated 3D-printed bone patches.}
  \begin{tabular}{lcc}
    \hline
    \textbf{Weight (g)} & \textbf{Sample 1} & \textbf{Sample 2} \\
    \hline
    \textbf{O} & 34.5 & 33.3 \\
    \textbf{F} & 29.1 & 28.8 \\
    \hline
  \end{tabular}

  \label{tab:sample_weights}
\end{table}

The sample masses, presented in Tab.~\ref{tab:sample_weights}, vary by up to 3\%. These variations were accounted for in the finite element models. For each sample, five measurement points on the surface were used for parameter identification. The mean and standard deviation of the identified parameters are presented in Fig.~\ref{fig:Phacon_head_patch}.
We observe that stiffness is consistently captured, ranging from 3.8–4 GPa at 100 Hz to 4.4–4.5 GPa at 5\,000 Hz. Damping remains relatively constant across all samples, within the range of 5–7\%. The broader confidence intervals at lower frequencies likely reflect the limited resolution inherent in linearly spaced frequency steps, which can reduce sensitivity to sharp resonances in this range. This appears to be a general limitation of the method rather than a result of measurement quality.

\section{Conclusion}
This study demonstrates the applicability of an inverse MOR approach for identifying frequency-dependent viscoelastic properties of structures with light damping. This damping level is particularly suitable because it preserves distinct peaks in the FRF, which are essential for accurately extracting dynamic characteristics. By using optical measurements of a single transfer function, the four parameters of the fractional derivative model can be accurately determined. \\
To validate this method, the storage modulus and loss factor of a POM beam were compared with DMA measurements, and the identification process is extended to a homogenized composite polymer-ceramic structure with curvature, showcasing varying levels of complexity.
In previous work, gradient-based optimization methods were suggested for similar inverse approaches. However, this is unrealistic when a broad range of material properties has to be considered. Uncertainty due to a mismatch between experimental and numerical data further limits the choice for an optimization algorithm. We conclude that in such cases global optimization schemes such as PSO are essential for converging to a global minimum. The PSO algorithm exhibits robust performance in solving this constrained optimization problem. 

Best results were obtained using measurements close to the force location, and having at least a number of five pronounced resonances within one FRF is needed for a robust fit. Overall, the 3D-printed structure highlights the inherent uncertainties associated with the material and manufacturing process. Factors such as printing speed, temperature, and material composition can affect mass distribution. Therefore, optimizing multiple points across multiple samples provides a reliable estimate of the properties of 3D-printed structures. Compared to DMA, this analysis represents a step forward, as very small 3D-printed samples are often too brittle and exhibit a non-homogeneous microstructure, making tests such as three-point bending impractical. The results also show that even over relatively small frequency ranges, the stiffness and damping vary up to 25\%. Assuming constant stiffness and damping values therefore leads to important deviations of the predicted resonance frequencies and amplitudes.

Future research may explore alternative constraint handling techniques or hybrid algorithms to enhance convergence speed and solution accuracy. Additionally, higher damping ratios and more complicated full-size structures would be interesting to cover a larger range of vibration problems. 

\section*{Acknowledgments}
This work was supported by grant 213127 of the Swiss National Science Foundation.
We thank Dr.\ Ivo Dobrev and Johannes Niermann (Department of Otorhinolaryngology, Head and Neck Surgery, University Hospital Zurich) for performing the micro-CT scanning and providing the microstructure image used in Fig.~\ref{fig:PhaconPatch}.

\section*{Competing interests} 
The authors have no competing interests to declare.

\appendix
\section{Sensitivity of modeled frequency response functions to the source and receiver position}

\begin{figure}[h!]
    \centering
    \begin{tikzpicture}
        \def\width{17cm}; % Define a fixed width for images

        % First image (BEM)
        \node[anchor=north] (img1) at (0,0) 
        {\includegraphics[width=\width]{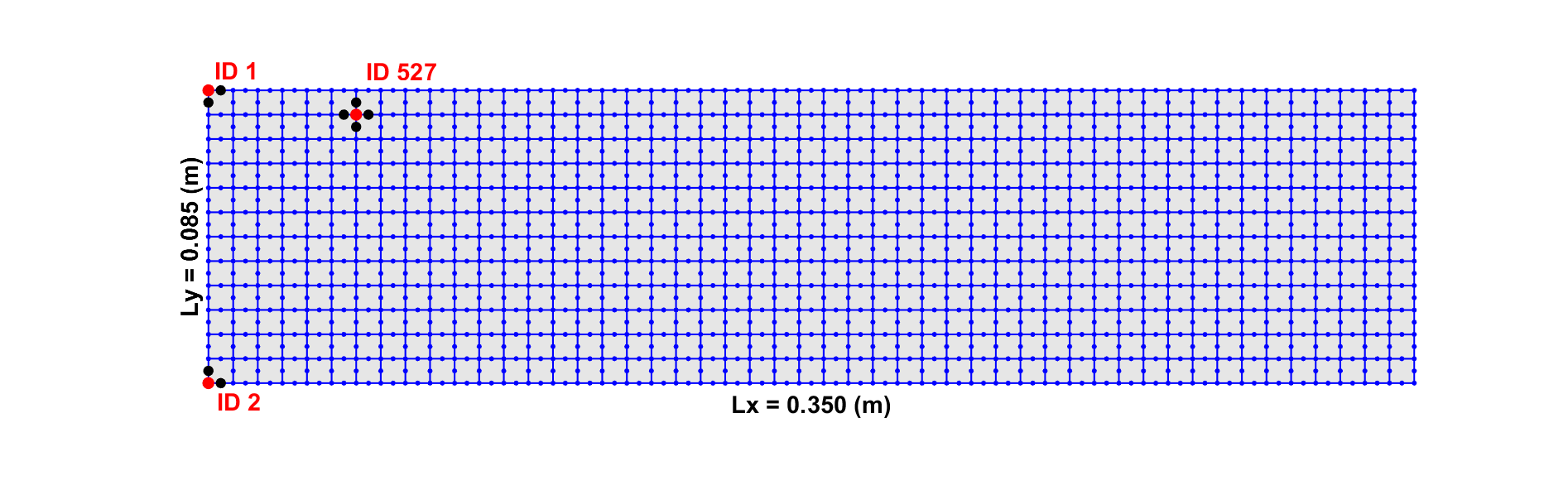}};

        % Second image (DCM) - Position below the first
        \node[anchor=north] at (img1.south) 
        {\includegraphics[width=\width,trim =0 25 0 25,clip]{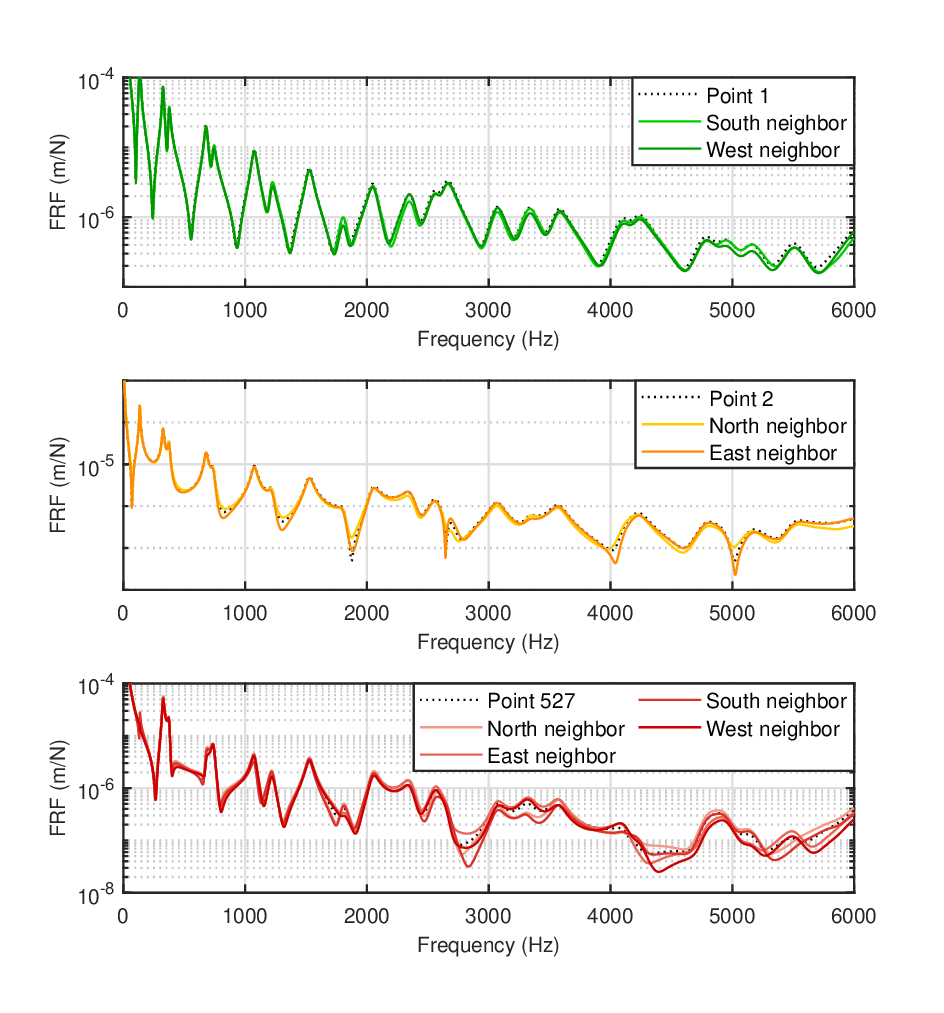}};
    \end{tikzpicture}
 \caption{\label{fig:point_sensitivity} Sensitivity of point-to-point FRFs for different measurement positions.}
\end{figure}

The choice of the measurement point influences the reconstructed viscoelastic properties. Since the identification process is based on a single measured point-to-point transfer function between the input force and surface velocity, understanding the sensitivity between measurement points is essential to assess potential uncertainty. For this reason, a finite element analysis is performed for one material model of the POM beam, as shown in Fig.~\ref{fig:point_sensitivity}. The quadratic elements have an edge size of 7mm. The structure is excited with a harmonic force applied to the top left corner element at the back of the structure. Three areas near the excitation source are selected for the FRF evaluation. 

Node number 1, which is closest to the force excitation, shows a FRF with pronounced resonances and minimal deviation from neighboring points located 3.5 mm away. Node number 2, situated on the opposite edge of the excitation direction, presents a more complex picture, since pronounced anti-resonances show slight deviations between the evaluation points. Since the vibration levels are close to the experimental noise floor, these deviations may lead to greater uncertainty. 

For node number 527, which is not located close to an edge of the beam, the  deviation between neighboring points becomes even more significant. The anti-resonances are more pronounced with larger differences. More importantly, this point lies near the node line of several bending eigenmodes of the structure, resulting in substantial deviations in both frequency and amplitude, e.g. at 2900 Hz and 4400 Hz.
These insights show that it is of utmost importance to match the evaluation position of the measured and modeled FRF. A measurement taken close to the force excitation appears to be more robust against small positional deviations.

\begin{figure}[h!]
    \centering
    \begin{tikzpicture}
        \def\width{17cm}; % Define a fixed width for images

        % First image (BEM)
        \node[anchor=north] (img1) at (0,0) 
        {\includegraphics[width=\width]{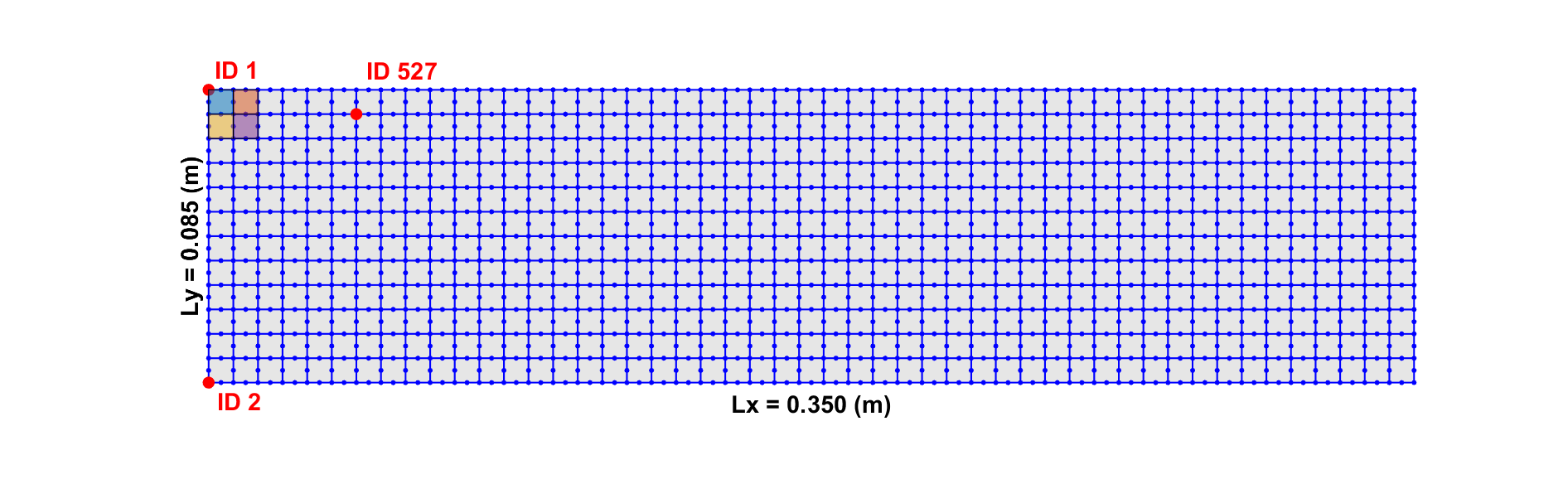}};

        % Second image (DCM) - Position below the first
        \node[anchor=north] at (img1.south) 
        {\includegraphics[width=\width,trim =0 20 0 25,clip]{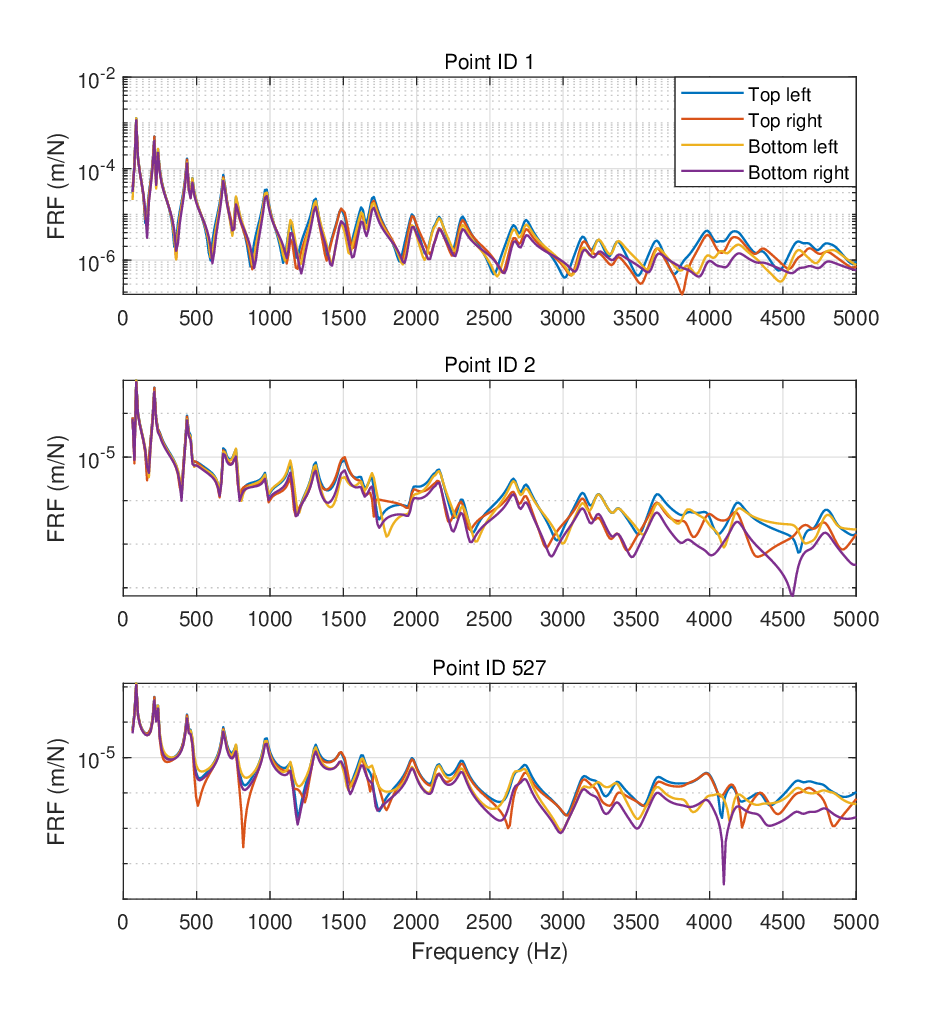}};
    \end{tikzpicture}
 \caption{\label{fig:point_sensitivity2} Sensitivity of point-to-point FRFs for different excitation positions.}
\end{figure}

In addition to the receiver position, the sensitivity of the excitation position needs to be investigated. Within this work, this excitation was always achieved by applying a surface force on a single element which has approximately the size of the permanent magnet, that was glued on the excited structure. It can be observed in the FRFs of the three illustrated points, that the excitation position even affects the points nearby significantly. Significant magnitude deviations become apparent also for points nearby the excitation location for frequencies higher than 1\,000 Hz, shown in Fig.~\ref{fig:point_sensitivity2}.

\newpage
% TODO
% Show FRF solutions: show accuarcy of basis

\bibliographystyle{unsrtnat}
\bibliography{reference.bib}

\end{document}